\documentclass[10pt,UKenglish]{article}
\pdfoutput=1
\usepackage[paper=a4paper, hscale=0.8]{geometry}
\usepackage{babel}
\usepackage{amsmath}
\numberwithin{equation}{section}
\usepackage{cite}
\usepackage{prettyref}
	\newcommand{\pr}[1]{\prettyref{#1}}
	\newrefformat{app}{appendix~\ref{#1}}
	\newrefformat{eqn}{eqn.~(\ref{#1})}
	\newrefformat{sec}{section~\ref{#1}}
	\newrefformat{subsec}{section~\ref{#1}}
	\newrefformat{subsubsec}{section~\ref{#1}}
	\newrefformat{fig}{fig.~\ref{#1}}
\usepackage{amsfonts}
\usepackage{dsfont}
\usepackage{graphicx}
\usepackage{xspace}
\usepackage{simplewick}
\usepackage{bbm}
\usepackage{url}
\newcommand{\im}[1]{\text{Im} \, [#1]}
\newcommand{\re}[1]{\text{Re} \, [#1]}
\newcommand{\ie}{i.e.,\xspace}
\newcommand{\nn}{\nonumber\\}
\newcommand{\landau}{\mathcal{O}}
\newcommand{\p}{\partial}
\newcommand{\eps}{\epsilon}
\newcommand{\V}[1]{\mathbf {#1}}
\newcommand{\dint}[2]{\int \frac {d^{#2} #1}{ (2 \pi)^{#2} } \,}
\newcommand{\vg}{\mbox{\boldmath$\nabla$}}
\newcommand{\bfmu}{\mbox{\boldmath$\mu$}}
\newcommand{\GeneralSelfEnergy}[2]{
\raisebox{#1}{
	\includegraphics[width=#2]{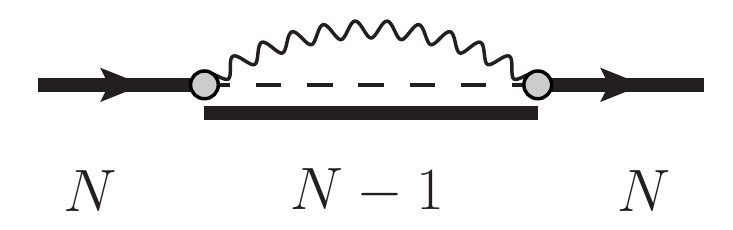}
}
}
\newcommand{\GeneralAtomPart}[2]{
\raisebox{#1}{
	\includegraphics[width=#2]{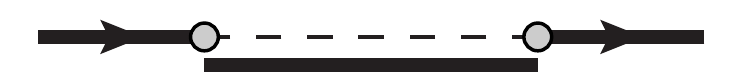}
}
}
\newcommand{\Vertexge}[2]{
\raisebox{#1}{
	\includegraphics[width=#2]{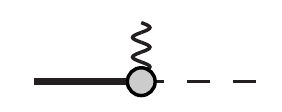}
}
}
\newcommand{\Vertexeg}[2]{
\raisebox{#1}{
	\includegraphics[width=#2]{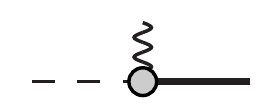}
}
}
\newcommand{\AtomPartBEC}[2]{
\raisebox{#1}{
	\includegraphics[width=#2]{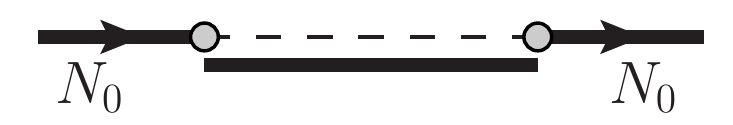}
}
}
\newcommand{\AtomPartIdeal}[2]{
\raisebox{#1}{
	\includegraphics[width=#2]{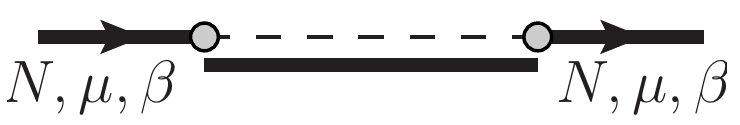}
}
}
\newcommand{\photon}[2]{
\raisebox{#1}{
	\includegraphics[width=#2]{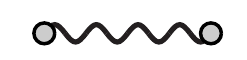}
}
}
\begin{document}
\renewcommand{\baselinestretch}{1.1}
\title{Interaction of a Bose-Einstein condensate with a surface:
perturbative S-matrix approach
}
\author{J\"urgen Schiefele  and Carsten Henkel
\\
\footnotesize
Universit\"at Potsdam,
Institut f\"ur Physik und Astronomie,
Karl-Liebknecht-Str. 24/25,
\\
\footnotesize
14\,476 Potsdam,
Germany
\\
\footnotesize
E-mail:
\url{Juergen.Schiefele@physik.uni-potsdam.de}
\normalsize
}
\date{}
\maketitle
%
%
%
%
%
%
%
%
%
%
%
%
%
%
%
%%%%%%%%%%%%%%%%%%%%%%%%%%%%%%%%%%%%%%%%%%%%%%%%%%%%%%%%%%%%%%%%%%%%%%%%%%%%%%
\abstract
%%%%%%%%%%%%%%%%%%%%%%%%%%%%%%%%%%%%%%%%%%%%%%%%%%%%%%%%%%%%%%%%%%%%%%%%%%%%%%
We derive an expression for the collective Casimir-Polder interaction 
of a trapped
gas of condensed bosons with a plane surface through the coupling
of the condensate atoms with the electromagnetic field.
% a $\bfmu . \V E$ 
% interaction of the condensate atoms with the electric field.
A systematic perturbation theory is developed based on a diagrammatic
expansion of the electromagnetic self-energy. 
In the leading order, the result for the interaction-energy is
proportional to the number of atoms in the condensate mode.
At this order, atom-atom interactions and recoil effects lead to
corrections compared to the single-atom theory, through shifts
of the atomic transition energies. We also discuss the impact of
the spatial delocalization of the condensate mode.
%
%
%
%
%
%
%%%%%%%%%%%%%%%%%%%%%%%%%%%%%%%%%%%%%%%%%%%%%%%%%%%%%%%%%%%%%%%%%%%%%%%%%%%%%%
\section{Introduction}
\label{sec:introduction}
%%%%%%%%%%%%%%%%%%%%%%%%%%%%%%%%%%%%%%%%%%%%%%%%%%%%%%%%%%%%%%%%%%%%%%%%%%%%%%
%
It is well known from cavity quantum electrodynamics (cavity QED) 
that the energy levels and lifetimes of the electronic states of an atom placed near a macroscopic body
are shifted from their free-space values \cite{Haroche92, Hinds94}.
This effect can be understood from
the modification the body imposes on the vacuum field modes which
lead, for example, to a position-dependent change in the Lamb shift. 
The resulting (van der Waals or Casimir-Polder) force between the atom and the macroscopic body has been shown
to match the predictions of QED in several experiments
\cite{Bender_2010,Sukenik_1993,Harber_2005,Landragin_1996,Obrecht_2007}.
% \todo{$\bullet$ Who (from many possibilities) should be quoted here?}
% everybody!

As cavity QED effects often do not require a relativistic treatment of the
electronic or atomic motion,
the techniques traditionally employed are lent from
non-relativistic QED:
a mode expansion of the electromagnetic field and a first quantized theory for 
the remaining (atomic) part of the system \cite{Barton_1988}.
Instead of working with mode expansions adapted to the presence 
of a body, there is another approach making use of the fluctuation-dissipation 
theorem \cite{Wylie_1984,Wylie_1985}:
the level shift is cast in a form involving generalized susceptibilities from linear
response theory, the (retarded) Green functions.
The influence of the surface is then encoded in the appropriate scattering 
amplitudes of the body, e.g., reflection coefficients for a planar interface.
This makes the approach applicable to very general descriptions of the
surface material, including absorption and dispersion.
Another advantage of the formalism lies in the fact that renormalization gets
simplified, as the (divergent) free-space part of the Lamb shift is
easily isolated from the surface-dependent contributions, the
latter being finite.

In the present paper, 
%we consider a condensed gas of interacting bosonic atoms 
%confined in a trapping potential above a surface.
we are interested in the shift of the collective energy levels of an 
$N$-atom system due to the presence of a nearby surface. 
A theory that has to account for the quantum statistical character of atoms
is conveniently formulated in terms of second-quantized
atom field operators.
We follow the standard procedure for perturbation theory,
which offers a pictorial representation in terms of Feynman diagrams
and permits us to calculate the elements of the electromagnetic self-energy,
approximating
them with the Dyson series \cite{Fetter}.
For our purposes, the theory has to deal with a confined atomic system in a trap
(including inter-atomic interactions),
and the interaction with the electromagnetic field is the relevant perturbation.
While much of the literature on Bose-Einstein condensates (BECs) 
in an external potential 
deals mainly with the collective properties of atoms in their 
electronic ground state,
a quantum field theory of ultra-cold atoms interacting with photons 
was formulated in
\cite{Lewenstein_1994, Walls_1994}.
We build on this approach and merge it with the linear response
techniques for electromagnetic field fluctuations near a surface.  
In the present paper, we consider the atom-light interaction up to second
order, which is the first non-vanishing contribution. We find under quite
general circumstances the atom-surface interaction energy and demonstrate
that it does not reduce to an integral over the density distribution of 
trapped (ground state) atoms. The propagation in the excited state,
although only virtually, connects ground state correlation functions at
different space-time points. This leads to a recoil shift of the atomic
polarizability, in addition to the familiar density shifts due to 
the atom-atom interaction. 
For the atomic ensemble, we consider two simple examples:
firstly, an interacting BEC at temperatures well below the critical 
temperature,
where we consider only a single mode of the atomic field with macroscopic  
occupation.
Our second example is the ideal Bose gas at nonzero temperature that can
be essentially characterized analytically.
Both systems are held in harmonic traps centered near the surface.
%and assume that $N_0$ ground state
%atoms populate the collective state of lowest energy,
%the condensate wave function.
We develop in this paper the main methods, check that several
limiting cases are recovered and discuss the two examples above in some
detail.
The aim for future publications is to generalize this approach in two respects:
on the cavity-{QED} side of the problem, to push the atom-field 
interaction to higher orders and, on the BEC side, to take into account 
low-lying collective states of the interacting atomic ensemble like 
Bogoliubov quasi-particles. 

The paper is organized as follows:
In \pr{sec:FToA}, we describe the interaction of the atomic system and
the electromagnetic field, in a form involving second quantized
operators for atoms as well as for the field.
In \pr{sec:S_matrix}, we calculate a general expression for the energy shift of the atomic ensemble
due to the interaction with the electromagnetic (e.m.) field
by evaluating the first non-vanishing term in the Dyson expansion of the 
$S$-matrix.
The result thus obtained 
involves the Feynman propagator for the e.m.\ field in the presence of a surface
which is introduced in \pr{subsec:E_sur}.
Atomic propagators 
are calculated in \pr{subsubsec:atom_part_BEC} and
\pr{subsubsec:atom_part_ideal}
for an interacting BEC (zero temperature)
and an ideal gas (nonzero temperature),
respectively.
The results of \pr{sec:FToA} to \pr{sec:propagators} are then used to 
calculate the atom-surface  
interaction of these two examples
(\pr{sec:BEC_har} and \pr{sec:ideal}).
%The influence of the surface is considered in \pr{subsec:E_sur}: 
%making use of the fluctuation-dissipation theorem, the appropriate retarded 
%Green functions 
%for the electric field are introduced into our general expression.
%Assuming a Gaussian shape for the condensate mode, we calculate an expression
%for the energy shift of the BEC trapped in front of a surface in \pr{sec:BEC_har},
%the details of the surface being modeled by appropriate reflection coefficients.
We cross-check our calculations against existing results
in \pr{subsec:single_atom}, by re-deriving the Casimir-Polder potential
for a single perfectly localized atom.

Our units are such that $\hbar = k_B = 1$, the speed of light
$c$ and the atomic mass $M$ are kept for the ease of reading.

%
%
%
%
%
%
%
%
%
%
%
%
%
%
%%%%%%%%%%%%%%%%%%%%%%%%%%%%%%%%%%%%%%%%%%%%%%%%%%%%%%%%%%%%%%%%%%%%%%%%%%%%%%
\section{Quantum field theory of atoms and photons}
\label{sec:FToA}
%%%%%%%%%%%%%%%%%%%%%%%%%%%%%%%%%%%%%%%%%%%%%%%%%%%%%%%%%%%%%%%%%%%%%%%%%%%%%%
%
We consider $N$ identical atoms in a trap above a flat surface.
The surface is taken to lie in the $xy$-plane, the center of the
trap is located a distance $d$ from the surface in the half-space $z > 0$.
The atoms are treated in the electric dipole approximation with an
electric ground state $|g\rangle$ and excited states $|e\rangle$.
The extension of this model to more realistic atoms is straightforward by summing
the contributions of all excited states in the calculation of the ground state shift.
%An extension to strong or long-range interactions among excited
%state atoms would be interesting for dipole or Rydberg blockade 
%problems~\cite{}.

Apart from possible inter-atomic interactions,
the atoms interact with the electromagnetic field via  a 
$\bfmu \cdot\V E$ interaction term,
where the dipole operator has transition matrix elements $\bfmu^{ge} =
\langle g | \V d | e \rangle$.
(For a comparison between the minimal coupling Hamiltonian and
$\bfmu \cdot\V E$ interaction, see \cite{Healy, Craig}.)
The interaction between the atomic system and the surface originates in 
this atom-field coupling:
the surface contains sources that radiate a field, and it
imposes boundary conditions on both the intrinsic field fluctuations and
the field radiated by the atom.
The relevant correlation functions of $\V E$ near the surface
will be dealt with in \pr{subsec:E_sur}. 

%\todo{
%$\bullet$ Suggestion for presentation: can we keep the formalism general
%as far as possible? This means: keep the option of both bosonic or fermionic 
%atom operators, make the `single-mode approximation' as late as possible.}
As in \cite{Lewenstein_1994}, we will work  with a Hamiltonian that describes
the atomic degrees of freedom (as well as the electric field) in second quantization
\ie a quantum field theory of atoms interacting with photons.
The operators $\Psi_g ( \V r)$ and $\Psi_e (\V r)$ describe the annihilation of an excited-state 
or ground state atom at location $\V r$.

As we want to treat the influence of the electromagnetic 
coupling as a perturbation
to the atomic system, we split
the total Hamiltonian as follows:
\begin{equation}
H = H_A + H_{AF} + H_{F}
\label{eqn:H0_Haf}
\end{equation}
Here, $H_{F}$ is the Hamiltonian for the unperturbed field
in the presence of the surface,
$H_{AF}$ contains the atom-field interaction,
and the Hamiltonian $H_A$ describes the trapped atoms.
The atomic operators in an  interaction-picture with respect to $H_{AF}$ then have
the general form
\begin{eqnarray}
\Psi_g (x)
	&=&
	\sum_{\V n} \,
	\Phi_{\V n} (\V r) \,
%	\operatorname{exp}[-i t E_g^{int} (\hat{N}_g,\hat{N}_e)] \,
	\hat{g}_{\V n}( t )
\label{eqn:Psi_g_general}
\;,
\\
\Psi_e (x)
	&=&
	\int \frac{d^3 q}{(2 \pi)^{3/2}} \,
%	\operatorname{exp}( i \V q . \V r ) - i t E_e^{int} (\hat{N}_g,\hat{N}_e)] \,
	\operatorname{exp}( i \V q . \V r )%
		% - i t E_e^{int} (\hat{N}_g,\hat{N}_e)] \,
	\hat{e}_{\V q}( t )
\;.
\label{eqn:Psi_e_general}
\end{eqnarray}
where the time dependence of the operators $\hat{g}_{\V n}( t )$ and
$\hat{e}_{\V q}( t )$ is specified in \pr{subsec:atom-propagator} below.
%
%The terms $E_g^{int}$ and $E_e^{int}$ are present if the unperturbed atomic 
% system features interaction between excited and ground state
% atoms or self-interaction amongst one or both species.
They satisfy
% anihilation and creation operators $\hat{g}_{\V n},\hat{g}^\dagger_{\V n}$ 
% and $\hat{e}_{\V q},\hat{e}^\dagger_{\V q}$
% are not time dependent and fulfill 
the bosonic or fermionic equal-time commutation relations.
In terms of these field operators, the atom-field interaction $H_{AF}$ in \pr{eqn:H0_Haf}
can be written as
%
%\todo{$\bullet$ look, this is quite general and independent of bosons/fermions
%and single-mode approximation}
%
\begin{equation}
H_{AF} =
	- \int d^3 x  \sum_\alpha
	\bigl\{
	 {E}_\alpha( x) \bigl[ \mu^{ge}_\alpha \Psi_g^\dagger ( x) \, \Psi_e ( x)  + \mu^{eg}_\alpha \Psi_e^\dagger ( x) \, \Psi_g ( x) \bigr]	 
	\bigr\}
\label{eqn:L_81}
\end{equation}
(compare \cite[eqn.~(81)]{Lewenstein_1994}).
As mentioned above, it is this term that is responsible for the interaction between the
surface and the atoms, as the specific form of $\V E (\V x)$ depends on the surface. We do not make the rotating wave (or resonance) approximation here
because otherwise relevant virtual processes would be missed. 
%We do neglect
%the possibility, of virtual atom pair creation from virtual photons. The latter
%would have energies so high that the electric dipole approximation would break
%down.
%

We use the notation 
$\V r = (\V x , z)$ for spatial vectors, 
where the two-dimensional vector $\V x$ lies in the plane perpendicular to the surface.
Spatial integrations $\int d^3 r$ run only over the $z > 0$ half-space.
Spacetime points are denoted by $x = (\V r , t)$.

%
%
%
%
%
%
%
%
%
%
%
%
%
%
%
%
%
%
%
%
%
%
%
%%%%%%%%%%%%%%%%%%%%%%%%%%%%%%%%%%%%%%%%%%%%%%%%%%%%%%%%%%%%%%%%%%%%%%%%%%%%%%
\section{Second-order energy shift}
\label{sec:S_matrix}
%%%%%%%%%%%%%%%%%%%%%%%%%%%%%%%%%%%%%%%%%%%%%%%%%%%%%%%%%%%%%%%%%%%%%%%%%%%%%%
%
The aim in this section is to calculate the energy shift
of the atomic system due to its interaction with the electric field.
In the case of a single atom in front of a surface,
this shift is usually calculated  in time-independent perturbation 
theory \cite{Wylie_1984,Wylie_1985,Gorza_2006}.
% As our many-atom system is described in terms of the field operators 
% $\Psi_g (x)$ and $\Psi_e (x)$, 
We will employ instead standard tools from field theory: 
the energy shift is obtained from the $S$-matrix, which can be perturbatively 
approximated with the Dyson series (see 
%\cite{Dyson_1949} or 
\cite[sec.~3.5]{WeinbergI}). %\footnote
{For a treatment of the single atom in front of a surface in both formalisms,
 nonrelativistic  perturbation theory and the Dyson series, 
 see \cite{Eberlein_2004, Eberlein_2006b}.
}

Let us briefly recall the basic relations which will be used:
the energy shift of an unperturbed state of the atomic system
can be calculated from the real part of the self-energy (logarithm of the
$S$-matrix). 
In the present paper, we will consider only terms up to 
the second order in $H_{AF}$ in which the self-energy
and the $T$-matrix coincide. Recall that 
the $T$-matrix is defined as the nontrivial part of the $S$-matrix,
% \todo{
% This approach works at lowest order. I think that we are actually 
% interested in the $log$ of the S-matrix. Recall that it scales like 
% $exp( - i T \delta H )$ where $\delta H$ is the change in the free
% Hamiltonian. Hence in the perturbation theory, we have to expand
% the $log$ of $S$ which gives, in fourth order, a mixing between second
% order and fourth order terms of the T-matrix. See the books on the
% self-energies and irreducible diagrams.}
%
\begin{equation}
S_{fi} = 
		\delta(f - i) 
		- 2 \pi i \, 
		\delta(E_f - E_i) \, 	
		T_{f i}
\; ,
\label{eqn:T_def}
\end{equation}
which, in turn, can be expressed as a series of time-ordered products of 
interaction picture operators, the Dyson series:
\begin{equation}
S = 	1 
	+ \sum_{n=1}^\infty \, 
	\frac{(-i)^n}{n!} \, 
	\int dt_1 \dots dt_n \, T\bigl\{ H_{AF}(t_1) \dots H_{AF}(t_n)  \bigr\}
\label{eqn:S_Dys}
\;,
\end{equation}
where the symbol $T\{ \dots \}$ denotes time ordering.

For a general self-interacting atomic system, it is convenient to define the interaction-picture operator
\begin{equation}
\Psi (x) 
	=
	\Psi_g^\dagger (x) 
	\Psi_e(x) 
\;,
%\label{eqn:}
\end{equation}
as the operators $\Psi_g$ and $\Psi_e$  appear only 
in this combination or its hermitian conjugate in \pr{eqn:L_81}.
With the initial and final states containing no 
excited-state atoms, the first-order
term in \pr{eqn:S_Dys} vanishes, leaving the second-order 
contribution
\begin{eqnarray}
S^{(2)}
        &=&
	\mu_\alpha^{eg} \mu_\beta^{ge} \,
	(-i)^2 \int d^4x_1 \, d^4x_2 \,
	\langle 
	\contraction{} {\Psi} {(x_2)} {\Psi}
	\Psi(x_2) \Psi^\dagger(x_1)
	\rangle \,
	D^F_{\alpha\beta}(x_2,x_1)
\label{eqn:S2N0b}
\\[1ex]
	&=&
	\GeneralSelfEnergy{-3.0ex}{0.15\textwidth}
\label{eqn:S2N0}
\; .
\end{eqnarray}
The brackets $\langle \dots \rangle$ in \pr{eqn:S2N0b} denote an expectation
value in a stationary state of the atomic Hamiltonian $H_A$.
In the above diagram, 
the in- and outgoing lines represent 
$N$ atoms in the state $|g\rangle$ that make up the unperturbed atomic 
state. The virtual state (inner line) 
consists of an atom in the state $|e\rangle$ (dashed line) 
propagating in the presence of a background field (solid line) made up of the
remaining $N - 1$ ground state atoms (still a large number).
The vertices, where an excited atom is created or destroyed,
are proportional to the dipole moment of the transition:
\begin{equation}
\Vertexge{-0.5ex}{0.07\textwidth} = -i \mu_\alpha^{eg}\;,
\qquad
\Vertexeg{-0.5ex}{0.07\textwidth} = -i \mu_\alpha^{ge}
\label{eqn:FR_vertex}
\;.
\end{equation}
The photon line in \pr{eqn:S2N0} is given by the time-ordered (or Feynman) 
propagator 
\begin{eqnarray}
%\photonline{-3.0mm} \quad
D_{\alpha\beta}^F (x_1 ; x_2) &=& 
	\photon{-0.25ex}{0.075\textwidth}
	=
	\langle \,  T \bigl\{ 
	E_\alpha(x_1) 
	E_\beta (x_2)  
	\bigr\} \, \rangle
\label{eqn:def_EF}
	\\
	&=& 
	\int \frac{d \omega}{2 \pi} \,
	e^{i \omega (t_1 - t_2)} \,
	\tilde{D}_{\alpha\beta}^F (\V {r}_1 , \V {r}_2; \omega) 
\label{eqn:def_EF_omega}	
	\;,
\end{eqnarray}
where the brackets $\langle \dots \rangle$ in \pr{eqn:def_EF} denote an expectation value with respect to 
an equilibrium state of $H_{F}$.

Finally, the contraction in \pr{eqn:S2N0b} 
%(see \cite{Wick_1950}  or \cite[chap.~3]{Fetter}) 
is defined as 
\begin{equation}
\contraction{} {\Psi} {(x_2)} {\Psi}
\Psi(x_2) \Psi^\dagger(x_1)
	=
	[\Psi(x_2), \Psi^\dagger(x_1)] \,
	\Theta (t_2 - t_1)
\;.
%\label{eqn:}
\end{equation}
which can be decomposed for bosonic or fermionic fields (upper/lower sign) as
\begin{align}
\contraction{} {\Psi} {(x_2)} {\Psi}
\Psi(x_2) \Psi^\dagger(x_1)
	&=
	\Theta (t_2 - t_1)
	\bigl\{
	\pm
	[\Psi_g^\dagger(x_2),\Psi_e^\dagger(x_1)]_\mp  \, [\Psi_e(x_2),\Psi_g(x_1)]_\mp 
\nonumber
\\	
	&+\;
	\Psi_g^\dagger(x_2) \, [\Psi_e(x_2),\Psi_e^\dagger(x_1)]_\mp \, \Psi_g(x_1) 
	\pm
	\Psi_e^\dagger(x_1) \, [\Psi_g^\dagger(x_2),\Psi_g(x_1)]_\mp \, \Psi_e(x_2)
\nonumber
\\
	&+\;
	[\Psi_g^\dagger(x_2),\Psi_e^\dagger(x_1)]_\mp \, \Psi_g(x_1)\Psi_e(x_2)
	\;+\; 
	\Psi_e^\dagger(x_1) \Psi_g^\dagger(x_2) [\Psi_e(x_2) , \Psi_g(x_1)]_\mp
	\bigr\}
\;.
\label{eqn:4com}
\end{align}
%
%\notes{10a_103}
If our initial and final states contain no excited atoms, the last three terms will 
yield zero in an expectation value,
and we are left with 
\begin{equation}
\langle 
\contraction{} {\Psi} {(x_2)} {\Psi}
\Psi(x_2) \Psi^\dagger(x_1)
\rangle
	=
	\langle
	\Psi_g^\dagger (x_2) \Psi_e (x_2) \Psi_e^\dagger (x_1) \Psi_g(x_1)
	\rangle
	\Theta(t_2 - t_1)
	=
	\GeneralAtomPart{-0.5ex}{0.15\textwidth}
\label{eqn:atom_part_general}
\end{equation}
%
%\notes{10a_104}
for both statistics.

We will see in \pr{eqn:atom_part_ideal_a} below that for an ideal gas,
% that if our unperturbed atomic system does not contain any interactions 
% between the excited and ground state
% atoms and no self-interaction amongst either excited or ground-state atoms, 
the above expression reduces
to the form that is usually obtained from applying Wick's theorem to a 
time-ordered product of four 
interaction picture operators (see \cite[chap.~3]{Fetter}). This is no longer
true in the general case (interacting atoms), and 
%a diagram like Eq.(\ref{eqn:atom_part_general})
%is not a proper Feynman graph: 
%the inner line represents an operator whose
%value depends on the in- and outgoing lines.
the Feynman-rules for translating a diagram containing a line like \pr{eqn:atom_part_general}
must take into account the presence of other lines due to the interaction with the background field.
%
%
%
%
%
%
%
%
%
%
%
%
%
%
%
%
%
%
%
%
%
%
%
%
%
%
%
%
%
%
%
%
%
%
%
%
%
%
%
%
%%%%%%%%%%%%%%%%%%%%%%%%%%%%%%%%%%%%%%%%%%%%%%%%%%%%%%%%%%%%%%%%%%%%%%%%%%%%%%
\section{Photon and atom propagators}
\label{sec:propagators}
%%%%%%%%%%%%%%%%%%%%%%%%%%%%%%%%%%%%%%%%%%%%%%%%%%%%%%%%%%%%%%%%%%%%%%%%%%%%%%
%
%\todo{$\bullet$ try to include the atom propagator as well here,
%giving as examples: condensate, bosons, fermions}
%
In order to evaluate the general expression \pr{eqn:S2N0b}, we now need to assume a concrete
form for the function $D^F_{\alpha\beta}$ of \pr{eqn:def_EF} 
--- the propagator of the electric field in the presence of a surface ---
and for the expression 
$
\langle
\Psi_g^\dagger (x_2) \Psi_e (x_2) \Psi_e^\dagger (x_1) \Psi_g(x_1)
\rangle
$
in \pr{eqn:atom_part_general}, characterizing the atomic ensemble.
For the former, we can rely largely on work presented in \cite{Wylie_1985},
which will allow us to apply our technique to very general surface materials.
Concerning the latter, we will focus on a pure condensate in a trap
(\pr{subsubsec:atom_part_BEC}) and on a trapped ideal Bose gas at nonzero
temperature (\pr{subsubsec:atom_part_ideal}).
%
%
%
%
%
%
%
%
%
%
%
%
%
%
%
%
%
%
%%%%%%%%%%%%%%%%%%%%%%%%%%%%%%%%%%%%%%%%%%%%%%%%%%%%%%%%%%%%%%%%%%%%%%%%%%%%%%
\subsection{Photon propagator near a surface}
\label{subsec:E_sur}
%%%%%%%%%%%%%%%%%%%%%%%%%%%%%%%%%%%%%%%%%%%%%%%%%%%%%%%%%%%%%%%%%%%%%%%%%%%%%%
%
%
The time-ordered propagator for the $\V E$-field [\pr{eqn:def_EF}]
is usually worked out explicitly from a mode expansion of the $\V E$-field.
This can be done in the presence of a non-dispersive surface, too, 
with the mode functions getting of course more cumbersome to satisfy
the boundary conditions at the surface
\cite{Mandel71, Eberlein_2006b}. 
We want to follow here the approach of 
\cite{Agarwal75a,Wylie_1984,Wylie_1985},
which connects the field propagator
to a form involving correlation functions from linear response theory,
the retarded Green functions. This applies as long as the fluctuation-dissipation
theorem for the electromagnetic field holds \cite{Callen51,Agarwal75a}.

The retarded Green function for the electric field is defined as  
\begin{eqnarray}
G_{\alpha\beta}(x_1, x_2) 
	&=&
	i \, \langle \,
 	[E_{\alpha}(x_1) , E_{\beta}(x_2)]   
	\, \rangle \, 
	\Theta(t_1-t_2)	
\label{eqn:G_ret}
%\\
%	&=&
%        i \, \operatorname{Tr} \,
%	\bigl\{
% 		e^{\beta H_{F}} \,
%		[E(x_1) , E(x_2)]   
%	\bigr\} \, 
%	\Theta(t_1-t_2)	
\end{eqnarray}
%
%(see \cite[sec.~32]{Fetter}) 
with Fourier transform
$G_{\alpha\beta}(\V{r}_1, \V{r}_2, \omega)$.
By rearranging the time ordered product in \pr{eqn:def_EF}
and using the fluctuation-dissipation theorem 
(see \cite[appendix~B]{Wylie_1985})
we can express the Fourier transform of the Feyn\-man-pro\-pa\-ga\-tor \pr{eqn:def_EF_omega} as
\begin{equation}
\tilde{D}^F_{\alpha\beta}(\V{r}_1 , \V{r}_2 , \omega) = 
	\im{G_{\alpha\beta} (\V{r}_1 , \V{r}_2 , \omega)} \, 
	\coth[\frac \omega {2 T } ]
	- i \,
	\re{G_{\alpha\beta} (\V{r}_1 , \V{r}_2 , \omega)}
\label{eqn:als_fr}
\end{equation}
(see \cite[sec.~31]{Fetter}), where $T$ is the temperature ($k_B = 1$)
of the field.
Here we assume the field and its sources in thermal equilibrium at the 
temperature $T_F$.
The atomic part of the system may have a different temperature and
is even allowed to be in a non-thermal state.
As we will see below, it is preferable to integrate the retarded 
Green function along the imaginary frequency axis.
Using the fact that $G_{\alpha\beta}(\omega)$ has only poles in the lower 
half of the
complex $\omega$-plane,
we can express the $\omega$-integration 
in $\langle N_0 | T^{(2)} | N_0 \rangle$ (see \pr{eqn:TN0} below) as
(see \cite[Appendix~A]{Gorza_2006}):
\begin{equation}
\int d\omega \, 
\frac	{\tilde{D}^F_{\alpha\beta}(\omega)}
	{\overline{\omega}_0  - \omega - i\eps} 
	\approx
	 2 \, \int_0^\infty d\xi \, 
	G_{\alpha\beta}(i\, \xi) \, 
	\frac{\overline{\omega}_0}{\overline{\omega}_0^2 + \xi^2}  
	\;+\;
	2 \pi \, G_{\alpha\beta}(-\overline{\omega}_0) 
	\Theta( -\overline{\omega}_0 )
%	\bigl( 
%		1 - \coth \bigl[ \frac{\overline{\omega}_0 }{ 2 T }  \bigr]  
%	\bigr) 
\;,
\label{eqn:f2rN0}
\end{equation}
where
$
\overline{\omega}_0 = \omega_{eg}({\bf q}, N_0)
$.
We have made here the approximation
$\overline{\omega}_0 \gg T_F$,
\ie field temperatures much smaller than the atomic transition energies,
where the summation over the poles of $\coth(\omega/2T)$ can
be replaced by an integral.
The second term on the rhs of \pr{eqn:f2rN0} is nonzero only for excited state atoms
($\overline{\omega}_0 < 0$) and describes spontaneous
emission and resonant contributions to the energy shift 
\cite{Hinds_1991, Gorza_2006}.
For atoms in the ground state, corrections to Eq.(\ref{eqn:f2rN0})
are proportional to the number of thermal photons which is exponentially 
small if $T_F$ is much smaller than the relevant transition frequencies.
% \ll \omega_{eg}( {\bf q}, N_0 )$.
When using \pr{eqn:f2rN0} in the remaining sections, we will suppose throughout that
the number of thermal photons is negligible, and any temperature dependence that appears from
now on is always associated with the temperature of the atoms, not the photon field.
The generalization to finite field temperatures is left for future work.

Now, from linear response theory (see \cite[sec.~32]{Fetter}) and the
linearity of the Maxwell equations, the response function
$G_{\alpha\beta}(\V{r}_1 , \V{r}_2 , \omega)$ can be identified with the 
classical Green function, 
\ie the electric field at $\V{r}_1$ generated by a classical dipole,
oscillating at frequency $\omega$,
which is located at $\V{r}_2$.
The explicit form of the Green function in the presence of an interface 
% ---and hence of $G_{\alpha\beta}(\V{x}_1 , \V{x}_2 , \omega)$---%
is well known \cite{Sipe_1981}
% which finally enables us to calculate the time-ordered
%propagator of the electric field via \pr{eqn:f2rN0}.
% The retarded Green function 
and can be split into a free space and a reflected part:
\begin{equation}
G_{\alpha\beta} = G^0_{\alpha\beta} + G^R_{\alpha\beta}
\label{eqn:gr_split}
\end{equation}
where $G^0_{\alpha\beta}$ is the retarded Green function in free space.
As we are only interested in that part of the energy shift caused by the presence of the surface,
we will not consider $G^0_{\alpha\beta}$ at all.
The decomposition \pr{eqn:gr_split} permits us in a simple manner 
to subtract the divergent diagrams involving photon loops that
yield the free-space Lamb shift, because the latter 
arise from the Green function $G^0_{\alpha\beta}$.
To get the distance-dependent part of the energy shift, 
we will simply substitute
$G_{\alpha\beta}$ by $G^R_{\alpha\beta}$. 
The expressions containing $G^R_{\alpha\beta}$ are then finite without 
any further renormalization. 
%
% \todo{Well, up to delicate issues as distance $d \to 0$.}

The surface contribution $G^R_{\alpha\beta}$ at imaginary frequencies
has the form % (we use $c = \varepsilon_0 = 1$)
%
% \todo{$\bullet$ check units for $\varepsilon_0$, sorry!
% $\to$ If $\varepsilon_0 = 1$, then our $G_{\alpha\beta}$ is the same as in 
% \cite[eqn.~(2.10)]{Wylie_1984}.}
%
\begin{equation}
G^R_{\alpha\beta}(\V {r}_1, \V{r}_2, i \xi) 
	= 
%	- \frac{\xi^2}{2\pi } \, 
	- \frac{\mu_0 \xi^2}{2\pi} \, 
	\int \frac{ d^2 k}{\kappa}
	R_{\alpha\beta}(\xi,\V k) 
	e^{-\kappa(z_1+z_2)} \, 
	e^{i \V k \cdot (\V{x}_1 - \V{x}_2)}
\;,
\label{eqn:G_def}
\end{equation}
(see \cite{Sipe_1981, Wylie_1984} and below in \pr{app:gr} for more details)
where $\mu_0 = (\varepsilon_0 c^2)^{-1}$ is the vacuum permeability
and 
$
\kappa = \sqrt{\xi^2 / c^2 + k^2}
.
$
The two-dimensional vectors
$\V x$ and $\V k$ denote position and momentum vectors 
parallel to the surface, respectively.
The tensor elements $R_{\alpha\beta}$ contain the reflection 
coefficients appropriate
for the specific surface material. 
As we are only considering the reflected part $G^R_{\alpha\beta}$, we will in the following skip the label `R' from \pr{eqn:G_def}.
Note that from the viewpoint of perturbation theory, the surface response
functions $R_{\alpha\beta}$ depend on the quantum state of matter
in the surface; they are calculated, of course, in the absence of the atomic
system outside it.
%
%
%
%
%
%
%
%
%
%
%
%
%
%
%
%
%
%
%
%
%
%
%%%%%%%%%%%%%%%%%%%%%%%%%%%%%%%%%%%%%%%%%%%%%%%%%%%%%%%%%%%%%%%%%%%%%%%%%%%%%%
\subsection{Propagators for atoms}
\label{subsec:atom-propagator}
%%%%%%%%%%%%%%%%%%%%%%%%%%%%%%%%%%%%%%%%%%%%%%%%%%%%%%%%%%%%%%%%%%%%%%%%%%%%%%
%
In the following, we calculate the atomic part of expression
\pr{eqn:S2N0b} for two simple examples of atomic systems.
Together with the photon propagator obtained in \pr{subsec:E_sur} above, these
will finally allow us to evaluate the atom-surface interaction in  \pr{sec:BEC_har}
and \pr{sec:ideal}.
%
%
%
%
%
%
%
%
%
%%%%%%%%%%%%%%%%%%%%%%%%%%%%%%%%%%%%%%%%%%%%%%%%%%%%%%%%%%%%%%%%%%%%%%%%%%%%%%
\subsubsection{Dilute interacting BEC in the single mode approximation}
\label{subsubsec:atom_part_BEC}
%%%%%%%%%%%%%%%%%%%%%%%%%%%%%%%%%%%%%%%%%%%%%%%%%%%%%%%%%%%%%%%%%%%%%%%%%%%%%%
%
For the interacting dilute Bose gas confined in a trap,
we further restrict ourselves to the deeply degenerate case, where
we can consider a large number $N_0$ of atoms in a single condensate mode.
We leave the contribution of condensate (Bogoliubov) excitations for future work.
The atomic Hamiltonian $H_A$ describes
two-level atoms with a contact interaction between excited and ground state atoms: 
%
%\todo{Check for $d^3 q/(2\pi)^3$. This depends on the commutation rules.}
%
\begin{equation}
H_A 	=  
	E(g_0^\dagger g_0) \, 
	+  
	\int d^3 q \,  
	\bigl( 
		\omega^e_{kin}(\V q) + b_{ge} g_0^\dagger g_0
	\bigr) 
	e^\dagger_{\V q} e_{\V q}
\label{eqn:HA_BEC}
\end{equation}
Here, the energy $\omega_e(\V q) = \omega_e + \V q^2 / 2M$ 
contains both the electronic excitation 
energy and the kinetic energy.
The constant $b_{ge}$ characterizes the interaction between ground- and excited state atoms.
The self-interaction amongst the ground state atoms and the effects of the trapping-potential are contained in
$E(g_0^\dagger g_0)$.
%Leaving the contribution of condensate (Bogoliubov) excitations 
%for future work, we focus here 
%on the collective ground state of the trapped atoms. 
%The atom number $N_0$ is thus an eigenvalue of the number operator 
% $g_0^\dagger g_0$.
In our approximation, the field operator $\Psi_g$ in \pr{eqn:Psi_g_general} 
consists only of a single mode
with the mode function $\phi_0 (\V r)$,
which is the condensate wavefunction calculated self-consistently by solving the Gross-Pitaevskii-equation (GPE) \cite{Stringari}
\begin{equation}
\bigl[  
	-\frac{\vg^2}{2M} + V_{tg}(\V r)  + b_{gg} \, (N_0 -1) \, |\phi_0(r)|^2  
\bigr] \, 
\phi_{0}(\V r) 
= 
\mu(N_0)  \, \phi_{0}(\V r) 
\label{eqn:GPE}
\; .
\end{equation}
%
%\todo{Check link between energy $E( \hat N_0 )$ and chemical potential
%$\mu( N_0 )$.}
%
Here, $\mu(N) = \p E(N)/\p N$ denotes the chemical potential,
%\notes{10B_102}
the constant $b_{gg}$ characterizes the self-interaction of ground-state atoms,
$V_{tg}$ denotes the trapping potential felt by the ground-state atoms,
and the condensate wavefunction is normalized to
$
\int d^3r \, |\phi_0(\V r)|^2 = 1
$.

Interactions between excited state atoms are neglected in this paper.
This is legitimate since our unperturbed state consists of a large
number of ground state atoms.
Excited state atoms will then only occur in virtual states, and their number
will be small. 
 
With the particular choice \pr{eqn:HA_BEC} for $H_A$, the field operators $\Psi_g$ and $\Psi_e$ from \pr{eqn:Psi_g_general} and \pr{eqn:Psi_e_general}
assume the form
\begin{eqnarray}
\Psi_g(x)
	&=& 
	\phi_0(\V r) \, 
	\operatorname{exp}[-i \, (E_g( \hat {N}_0 + 1) - E_g( \hat {N}_0 ) + b_{ge} \hat{N}_e) \, t] \, 
	\hat {g}_0  
\label{eqn:Psi_g_BEC}
\;,
\\
\Psi_e(x) 
	&=& 
	\int \frac{d^3 q}{(2 \pi)^{3/2}} \, 
	\operatorname{exp}[i \,(\V q . \V r  - ( \omega^e_{kin}(\V q) + b_{ge} \, \hat{N}_0 ) \, t )]\, 
	\hat{e}_{\V q}
\;.
\label{eqn:Psi_e_BEC}
\end{eqnarray}
%
%\notes{09e_43}
The diagram \pr{eqn:atom_part_general} then becomes
\begin{eqnarray}
\AtomPartBEC{-0.5ex}{0.15\textwidth}
%\langle
%\Psi_g^\dagger (x_2) \Psi_e (x_2) \Psi_e^\dagger (x_1) \Psi_g(x_1)
%\rangle
%\Theta(t_2 - t_2)
	&=&
	\Theta (t_2 - t_1) \,
	\phi_0(\V{r}_1) \phi_0^* (\V{r}_2) \, 
	N_0 \,
	\int \frac {d^3 q} {(2 \pi)^3} \,
	e^{i \V{q} . ( \V {r}_2 - \V {r}_1)} 
\nonumber\\
	&&\times \,
	e^{-i \omega_{eg}(\V q, N_0)	
	(t_2 - t_1)}
\;.
\label{eqn:atom_part_BEC}
\end{eqnarray}
Here the transition frequency $\omega_{eg}(\V q, N)$ 
is defined as
\begin{equation}
\omega_{eg}(\V q, N) = \omega_e( \V q ) 
+ E_g(N-1) - E_g(N) +   (N-1) b_{ge}
\label{eqn:w_BEC} 
\;,
\end{equation}
where the frequency shift of the atomic transition due to inter-atomic 
interactions appears.
If the  system consists of a single atom only,
\begin{equation}
\omega_{eg}(\V q, 1) = \omega_e( \V q )
- \omega_g = \omega_{eg}  + \frac{ \V q^2 }{ 2 M }
\label{eqn:w_BEC_limit}
\;,
\end{equation}
which is the resonance frequency of a single atom, including the recoil shift.
The physical interpretation of Eq.(\ref{eqn:atom_part_BEC}) is quite clear:
a virtual photon takes a ground state atom at position ${\bf r}_1$
to the excited state, the atom propagates freely to position ${\bf r}_2$
and joins the other ground state atoms there. We shall see below that
the relevant distances $| {\bf r}_2 - {\bf r}_1 |$ are negligibly small so
that eventually the ground-state density $|\phi_0(\V{r}_1)|^2$ determines
the atom-surface interaction.
%
%
%
%
%
%
%
%
%
%
%%%%%%%%%%%%%%%%%%%%%%%%%%%%%%%%%%%%%%%%%%%%%%%%%%%%%%%%%%%%%%%%%%%%%%%%%%%%%%
\subsubsection{Ideal {B}ose gas at finite temperature}
\label{subsubsec:atom_part_ideal}
%%%%%%%%%%%%%%%%%%%%%%%%%%%%%%%%%%%%%%%%%%%%%%%%%%%%%%%%%%%%%%%%%%%%%%%%%%%%%%
%
For the non-interacting trapped Bose gas,
treated in the grand-canonical ensemble with a mean total particle number $N$, 
an inverse temperature $\beta$ and chemical potential $\mu$, 
the Hamiltonian $H_A$ takes the form
\begin{equation}
H_A
	=
	\sum_{\V n} \, E_{\V n}
	+
	\int d^3 q \, \omega^e_{kin}(\V q) \hat{e}^\dagger_{\V q} \hat{e}_{\V q}
\;.
\label{eqn:HA_ideal}
\end{equation}
The mode functions for the operator $\Psi_g$ are the single-particle wavefunctions $\phi_{\V n}$
that solve
\begin{equation}
\bigl[  
	-\frac{\vg^2}{2M} + V_{tg}(\V r) 
\bigr] \, 
\phi_{\V n}(\V x) 
= 
E_{\V n}  \, \phi_{\V n}(\V r) 
%\label{eqn:}
\; .
\end{equation}
The field operators \pr{eqn:Psi_g_general} and \pr{eqn:Psi_e_general} now take the simple form
\begin{eqnarray}
\Psi_g(x)
	&=&
	\sum_{\V n} \, \phi_{\V n} (\V r) 
	e^{-i E_{\V n} t}
	\hat{g}_{\V n}
\label{eqn:Psi_g_ideal}
\;,
\\
\Psi_e(x)
	&=&
	\int \frac{d^3 q}{(2 \pi)^{3/2}}\,
	e^{i (\V q . \V r - \omega_{kin}^e(q) t)}
	\hat{e}_{\V q}
\;.
\label{eqn:Psi_e_ideal}
\end{eqnarray}
%
%\notes{09f_38,9}
The atomic part \pr{eqn:atom_part_general} yields
\begin{eqnarray}
\AtomPartIdeal{-0.5ex}{0.15\textwidth}
%\langle N|
%\contraction{} {\Psi} {(x_2)} {\Psi}
%\Psi(x_2) \Psi^\dagger(x_1)
%| N \rangle \,
	&=&
	\langle \Psi_g^\dagger (x_2) \Psi_g (x_1) \rangle \,
	\contraction{}{\Psi_e}{(x_2)}{\Psi_e^\dagger}
	\Psi_e(x_2)\Psi_e^\dagger(x_1)
\label{eqn:atom_part_ideal_a}
\\
	&=&
	\Theta(t_2 - t_1) \,
%	\langle \Psi_g^\dagger (x_2) \Psi_g (x_1) \rangle
	\sum_{\V n} 
	\phi_{\V n}^*(\V r_2) \phi_{\V n} (\V r_1)
  	{\rm e}^{ i E_{\V n} (t_2 - t_1)}
	\langle \hat{g}_{\V n}^\dag \hat{g}_{\V n}^{\phantom\dag} \rangle
%	\langle \Psi_g^\dagger (x_2) \Psi_g (x_1) \rangle
\label{eqn:atom_part_ideal}
\\
	&&\times
	\int \frac{d^3 q}{(2 \pi)^3} \,
	\operatorname{exp}[i \V{q}.(\V{r_2} - \V{r_1}) 
	-i (q^2/(2m) + \omega_{eg}) (t_2 - t_1)]
\nonumber
\end{eqnarray}
%
%\notes{10a_84,100}
%
Note again the occurrence of the two-point correlation
function for the ground-state atoms. 
We thus reach a similar structure 
as in Eq.(\ref{eqn:atom_part_BEC}) above,
but with a sum over all trap eigenstates.
%
%
%
%
%
%
%
%
%
%
%
%
%
%
%
%
%
%
%
%
%
%
%
%
%
%
%
%
%
%%%%%%%%%%%%%%%%%%%%%%%%%%%%%%%%%%%%%%%%%%%%%%%%%%%%%%%%%%%%%%%%%%%%%%%%%%%%%%
\section{Energy shift of an interacting {B}ose gas trapped near a surface}
\label{sec:BEC_har}
%%%%%%%%%%%%%%%%%%%%%%%%%%%%%%%%%%%%%%%%%%%%%%%%%%%%%%%%%%%%%%%%%%%%%%%%%%%%%%
%

\subsection{Generalized polarizability}

With the results obtained above, we can now evaluate the interaction potential
between the single-mode condensate and a surface.
Putting the expression for the photon propagator \pr{eqn:def_EF} and for the 
atomic two-point function \pr{eqn:atom_part_BEC} 
into \pr{eqn:S2N0b}, 
we get for the $T$-matrix element
(after performing the $dt_1$ and $dt_2$ integrations)
\begin{eqnarray}
\langle N_0 | T^{(2)} | N_0 \rangle &=& 
	N_0  \, \mu_\alpha^{ge} \mu_\beta^{eg} \, 
	\int d^3{r}_1 \int d^3{r}_2 \, \phi_0(\V{r}_2) \phi_0^*(\V {r}_1)
\nn
&&\times
	\int \frac{d \omega}{2\pi} 
	\, \tilde{D}^F_{\alpha\beta}(\V {r}_1 , \V{r}_2, \omega)
	\dint q 3 \frac{e^{i \V q \cdot (\V {r}_1 - \V{r}_2)}} 
	{\omega - \omega_{eg}(\V q, N_0) + i\eps }
\;,
\label{eqn:TN0}
\end{eqnarray}
%
%\material{The $dt_{1,2}$ integrals are done by using
%%
%\begin{equation}
%\int dt_1 \int dt_2 \, 
%\int\frac{d \omega}{2 \pi} \,
%\operatorname{exp}
%	\bigl[
%	- i (t_2 -t_1) (\omega_{eg} - \omega)
%	\bigr]
%\Theta(t_2 -t_1)
%	=
%	2 \pi i\,
%	\delta(0)\,
%	\int\frac{d \omega}{2 \pi} \,
%	\frac 1 {\omega -\omega_{eg} + i \epsilon}
%%\label{eqn:}
%\end{equation}
%%
%\notes{10a_20}
%.}
Using relation \pr{eqn:f2rN0} in \pr{eqn:TN0} to link 
the time-ordered photon propagator to the Green tensor $G_{\alpha\beta}$,
we get
\begin{eqnarray}
\langle N_0 | T^{(2)} | N_0 \rangle &=& 
	- N_0 \, \frac 2 {(2\pi)^4}  \,  
	\mu_\alpha^{ge} \mu_\beta^{eg} \, 
\int d^3{r}_1 \! \int d^3{r}_2 \,
	\phi_0(\V{r}_2) \phi_0^*(\V {r}_1) \,
\nn
&&\times
	\int_0^\infty d\xi \,
	G_{\alpha\beta}(\V{r}_1 , \V{r}_2 , i \xi)
	\int d^3 q  \, a({\V q}, \xi) \,
	e^{i {\V q} \cdot ( {\V r}_1 -{\V r}_2 )} 
\;,
\label{eqn:TN0.1}
\end{eqnarray}
The generalized polarizability
\begin{equation}
a({\V q}, \xi) 
	=
	\frac{\omega_{eg}( {\bf q}, N_0 ) }{ 
	\omega_{eg}^2( {\bf q}, N_0 )  + \xi^2}
\label{eqn:a_def}
\;.
\end{equation}
contains the interaction- and recoil-shifted resonance frequency
$\omega_{eg}( {\bf q}, N_0 )$ (see \pr{eqn:w_BEC}).
In \pr{eqn:TN0.1} we neglected the resonant contribution of 
thermally excited photons.

\subsection{Condensate wave function}

In order to evaluate \pr{eqn:TN0.1}, we have to substitute a suitable 
approximation
for the condensate wave function $\phi_0(\V r)$.
For simplicity we 
solve the Gross-Pitaevskii equation \pr{eqn:GPE}
with an isotropic harmonic trapping potential
\begin{equation}
V_{tg}(\V r) = \frac M 2 \, \nu^2 \, (\V{x}^2 + (z-d)^2) 
\label{eqn:V_def}
\;,
\end{equation}
where  $d$ denotes the distance of the trap center from the surface. 
If the kinetic term in the GPE can be neglected (Thomas-Fermi approximation),
the solution for the density profile takes the form of an inverted parabola.
This is usually a good approximation for large particle numbers.
Here, we choose a Gaussian ansatz for the wave function because it simplifies
the subsequent integrations. (For calculations with a Thomas-Fermi profile,
see \cite{Klimchitskaya08c}.)
The ansatz also allows for the limit $N_0 \to 1$ 
in order to provide a cross-check with results for a single-atom system (\pr{subsec:single_atom}).
Gaussian functions also approximately solve the GPE, 
if width and amplitude are varied such that the Gross-Pitaevskii functional is minimized 
(see 
\cite{Stringari}
%\cite{Lewenstein_Zoller_1997} 
for details).
We thus make the ansatz
\begin{equation}
\phi_0 (\V r) = 
	(\sqrt{\pi} \, \sigma(N_0))^{- \frac 3 2} \, 
	\operatorname{exp} \left[- \frac{ \V{x}^2  + (z-d)^2 
	}{ 2 \sigma^2(N_0) }  
 \right]
\label{eqn:wfct}
\;.
\end{equation}
The minimization procedure gives a spatial width $\sigma( N_0 )$ 
in \pr{eqn:wfct} that depends on the number of trapped particles
and has the asymptotic values~\cite{Lewenstein_Zoller_1997}
%%
%\begin{equation}
%\sigma (N_0) = a_0 \, \overline{\sigma}(N_0)
%\label{eqn:sigma_def}
%\;,
%\end{equation}
%
\begin{equation}
\sigma( N_0 ) =
\begin{cases}
a_0  ,&  N_0=1
\\
a_0
\biggl( \displaystyle
\sqrt{\frac 2 \pi} 
\frac{N_0 a}{a_0} \biggr)^{1/5} 
,&  
\displaystyle
\frac{N_0 a}{a_0} \gg 1
\end{cases}
\;.
	\label{eqn:gaussian-bec-width}
\end{equation}
where $a_0 = (M \, \nu)^{-1/2}$ is the width of the single-particle ground
state in the trap. The $s$-wave scattering length $a$
is related to the interaction constant $b_{gg}$ from \pr{eqn:GPE} 
via
$
b_{gg} = 4 \pi a / M
$.  
In the second case of \pr{eqn:gaussian-bec-width}, the interaction energy 
of ground state atoms 
is much larger than the bare harmonic potential.
This regime corresponds to the Thomas-Fermi limit (the Thomas-Fermi
radius is $R_{\rm TF} = a_0 (15 N_0 a / a_0)^{1/5}$).

A subtlety arises for the Gaussian ansatz~(\ref{eqn:wfct}) because 
it is normalized only in the limit $d \gg \sigma(N_0)$ if
spatial integrations are restricted over the half-space $z > 0$.
We shall always assume this limit, %$d \gg \sigma(N_0)$, 
as our approach is clearly not valid for atoms touching the surface.
The wave function $\phi_0$ is of the 
order $\landau(\operatorname{exp}[- (d / \sigma(N_0))^2])$
at the surface, and exponentially small terms of this order will be systematically
discarded in numerical evaluations of energy shifts 
in \pr{subsec:single_atom} and \pr{sec:ideal}.
These approximations are dealt with in detail in \pr{app:erf}.

In the following, we will evaluate $T^{(2)}_{N_0 N_0}$ with the approximate
ground-state wave function $\phi_0$ from \pr{eqn:wfct}.
The result \pr{eqn:TN0.1} for the $T$-matrix is, however, more generally
valid and can be evaluated similarly for 
other approximations of $\phi_0(\V r)$.

\subsection{Recoil shift and (de)localization correction}
\label{subsec:recoil-correction}

At this stage, it is convenient to introduce sum and difference coordinates
${\V r}_\pm$ and to split them in components perpendicular and parallel 
to the surface:
$
z_\pm = z_1 \pm z_2
$,
and
$
\V {x}_\pm = \V{x}_1 \pm \V{x}_2
$,
Similarly, for the momentum, we use from now on
$
\V q = (q_x,q_y,q_z) \to (\V q, q_z)
$.
Integrating in \pr{eqn:TN0.1}
over $\V{r}_1$ and $\V{r}_2$ and the angle of the two-dimensional 
vectors $\V k$ and $\V q$ 
($\V k \cdot \V k = k^2$ and $\V q \cdot \V q = q^2$), we get
\begin{eqnarray}
\langle N_0 | T^{(2)} | N_0 \rangle &=& 
	- N_0 \,  \frac {2 \sigma^3} {\pi^{3/2}}  \,  
	\frac{ \mu_\alpha^{ge} \mu_\beta^{eg} }{ \varepsilon_0 } \, 
	\int_0^\infty d\xi \int_0^\infty \frac{k \, dk}{\kappa} \,
	\frac 1 2 \, e^{- 2 \kappa d } \, e^{\kappa^2 \sigma^2} 
	(1 + \operatorname{erf}[\frac d {\sigma} - \kappa \sigma])  \,
	M_{\alpha\beta}
\nn
&&\times
	\int_0^\infty d q_z \int_0^\infty q \, dq \, 
	a(q, q_z, \xi) \, 
	I_0[2 \sigma^2 k q] \,	
	e^{- (k^2 + q^2 + q_z^2) \sigma^2 } \,
\;,
\label{eqn:TN0.1a}
\end{eqnarray}
with an obvious notation for $a( q, q_z, \xi )$.
The diagonal matrix $M_{\alpha\beta}$ originates from the scattering
tensor $R_{\alpha\beta}$ and has elements
\begin{align}
M_{xx}(k,\xi) &= M_{yy}(k,\xi) = R^p \, k^2 + (R^p - R^s) \,  (\xi/c)^2 \;, 
\\
M_{zz}(k,\xi) &= 2 R^p \, k^2
\label{eqn:M_def}
\;,
\end{align}
and the $R^{s,p}( k, \xi )$ are the reflection amplitudes from the surface 
(eqns. (\ref{eqn:Rs_def}) and (\ref{eqn:Rp_def})),
$I_0$ is a Bessel function of the second kind, 
$\kappa = \sqrt{k^2 + \xi^2}$, and $\operatorname{erf}$ denotes the 
error function.

To perform the $dq_z$ and $dq$-integrations in \pr {eqn:TN0.1a},
we observe that in $a(q, q_z, \xi)$ (see \pr{eqn:a_def}) the
momenta $q$ and $q_z$ appear only as recoil shifts of the atomic
transition frequency $\omega_{eg}( N_0 )$ (see \pr{eqn:w_BEC}). 
Since the relevant momenta
are limited to typically $1/\sigma$, the recoil shift is a small correction
because $1/(M \sigma^2) = \nu \ll \omega_{eg}( N_0 )$ is usually well 
satisfied. 
We therefore expand
% Lamb-Dicke limit \sigma << \lambda_A
% here, 1/\sigma^2M << \omega_{eg}
% or \sigma << 1/(M \omega_{eg})^{1/2}
in powers of $q$ and $q_z$
and integrate term by term by 
means of the identities  
\begin{eqnarray}
\int_0^\infty q \, dq \, e^{- (k^2 + q^2) \sigma^2 } \, I_0[2 k q \sigma^2 ]
    &=&
    \frac 1 {2 \sigma^2}
\;,
\\
\int_0^\infty q \,  dq \, q^2 \, e^{- (k^2 + q^2) \sigma^2 } \, 
I_0[2 k q \sigma^2 ]
    &=&
    \frac 1 {2 \sigma^2} \big(\frac 1 {\sigma^2} +  k^2\big)
\;.
\end{eqnarray}
Thus, we finally obtain for the $T$-matrix \pr{eqn:TN0.1a} 
\begin{eqnarray}
\langle N_0 | T^{(2)} | N_0 \rangle &=& 
    -\frac {N_0} {2 \pi} \, 
	\frac{ \mu_\alpha^{ge} \mu_\beta^{eg} }{ \varepsilon_0 } 
    \int_0^\infty d\xi \, 
    \int_0^\infty \frac{k \, dk}{\kappa} \,
    \frac 1 2 \, e^{- 2 \kappa d } \, e^{\kappa^2 \sigma^2} (1 + \operatorname{erf}[\frac d {\sigma} - \kappa \sigma])   \,
    M_{\alpha\beta}
\nn
&&\times
    \bigl\{
    \alpha(\xi,N_0)
    +
    \alpha^{(rc)}(\xi,N_0,k) 
    \bigr\}
\label{eqn:TN0.3}
\end{eqnarray}
where the polarizability
\begin{equation}
\alpha(\xi,N_0)
    =
    \frac{\omega_{eg}(N_0)}{\omega_{eg}(N_0)^2 + \xi^2} 
\;,
\label{eqn:alpha_def}
\end{equation}
describes the no-recoil case. 
The recoil term $\alpha^{(rc)}$ is given by
\begin{equation}
\alpha^{(rc)}(\xi,N_0,k) 
    =
    - \frac{\omega_{eg}(N_0)^2  - \xi^2} {(\omega_{eg}(N_0)^2  + \xi^2)^2} \,
    \biggl( 
	    \frac{3}{4 M (\sigma(N_0))^2}  
	    + \frac{k^2}{2 M} 
 	\biggr) 
\;.
\end{equation}
We can attribute
this correction % in \pr{eqn:TN0.3} % can  be reformulated   
% as an energy shift equivalent 
to a recoil shift of the effective resonance frequency
\begin{equation}
\omega_{eg} \to 
    \omega_{eg} 
    + 
%    \biggl( 
	    \frac{3}{4 M (\sigma(N_0))^2}  
	    + \frac{k^2}{2 M} 
%    \biggr) 
\end{equation}
where the two terms describe the kinetic energy from the delocalized 
condensate wave function and from the absorbed photon momentum 
in the excited state, respectively.
%
%We recognize in $T^{(2)\,rc}_{N_0 N_0}$ the first order expansion of
%the polarizability~(\ref{eqn:alpha_def}) as the resonance frequency
%$\omega_{eg}( N_0 )$ is shifted,
%%
%\begin{equation}
%\frac{\omega_0 - x}{(\omega_0 - x)^2 + \xi^2} = 
%	\frac{\omega_0}{\omega_0^2 + \xi^2}
%	+ x \, \frac{\omega_0^2 - \xi^2}{(\omega_0^2 + \xi^2)^2}
%	+ \landau(x^2)
%\end{equation}
%%
% for small $x$, 

The T-matrix element
$T^{(2)}_{N_0 N_0}$ from \pr{eqn:TN0.3} is our main result for the 
interaction 
energy of a trapped Bose gas with a plane surface.
In the above form, it is clear that $T^{(2)}_{N_0 N_0}$ generalizes 
the result  for a stationary single atom 
in a straightforward manner.
Clearly, as we put $N_0=1$, we get
the single-atom transition frequency
$\omega_{eg}(1) = \omega_{eg}$. 
And with the identity
\begin{equation}
\lim_{\sigma \to 0}
%    \biggl[
    \frac 1 2 \, e^{- 2 \kappa d } \, e^{\kappa^2 \sigma^2} (1 + \operatorname{erf}[\frac d {\sigma} - \kappa \sigma]) 
%    \biggr]
    =
    e^{- 2 \kappa d}
\;,
\end{equation}
we get from the no-recoil term of \pr{eqn:TN0.3}
\begin{equation}
\lim_{\sigma \to 0} 
% \bigl[ 
T^{(2)}_{1 1} 
% \bigr] 
    =
    -\frac {1} {\pi} \, 
    \mu_\alpha^{ge} \mu_\beta^{eg} \,
    \int_0^\infty d\xi \,
    G^R_{\alpha\beta}({\V r}_0 ,{\V r}_0 , i\xi ) \,
    \frac{\omega_{eg}}{\omega_{eg}^2 + \xi^2}	 
\label{eqn:cc_wylie}
\;
\end{equation}
with $\V{r_0} = (0,0,d)$ the position of the trap center.
This is the known result for a perfectly localized single atom as in
Ref.\cite[eqn.~(2.28)]{Wylie_1985}. The recoil correction 
involving $\alpha^{(rc)}$ is discussed
in more detail in \pr{subsec:single_atom}.
It is usually very small, unless the trap frequency $\nu$ is comparable to 
the atomic resonance $\omega_{eg}$, a case of no practical significance. 

For a large atom number $N_0$, the resonance frequency 
$\omega_{eg}(N_0)$ in \pr{eqn:alpha_def} incorporates the 
inter-atomic
interactions (see \pr{eqn:w_BEC}).
The overall proportionality factor $N_0$ of \pr{eqn:TN0.3} can be understood
by recalling that the responsible diagram (see \pr{eqn:S2N0})
represents a sum of  self-energies of $N_0$ individual ground state atoms.
In higher orders, \ie diagrams with four or more vertexes,
virtual photons can connect different ground state atoms,
and we can expect a nonlinear scaling in $N_0$.
%
%
%
%
%
%
%
%
%
%
%
%
%
%
%
%%%%%%%%%%%%%%%%%%%%%%%%%%%%%%%%%%%%%%%%%%%%%%%%%%%%%%%%%%%%%%%%%%%%%%%%%%%%%%
\subsection{Single ground-state atom}
\label{subsec:single_atom}
%%%%%%%%%%%%%%%%%%%%%%%%%%%%%%%%%%%%%%%%%%%%%%%%%%%%%%%%%%%%%%%%%%%%%%%%%%%%%%
%
The $T$-matrix element for a single atom can be obtained 
from \pr{eqn:TN0.3}
by setting $N_0 = 1$.
% $\omega_{eg}(N_0) \to \omega_{eg}$, $\sigma(N_0) \to a_0$
% and $1/M \to \nu a_0^2$.
Introducing the scaled distance $x = d \omega_{eg} / c$,
and rescaling the integration variables
$\overline{\xi} = \xi/\omega_{eg}$,
$\overline{k} = c k/\omega_{eg}$,
$\overline{\kappa} = c \kappa/\omega_{eg}$,
%
%\todo{$\bullet$ scale to free space linewidth which is a convenient energy
%unit: $\gamma_{eg} = |{\bf d}^{eg}|^2 \omega_{eg}^3 / 3\pi$.}
%
%\todo{$\bullet$ mention scaling of integration variables}
%
the $T$-matrix reads
%\notes{10b_46}
\begin{eqnarray}
\langle 1 | T^{(2)} | 1 \rangle
    &=& 
    -
%\frac{ 3 \gamma_{eg} }{ 2 }
    	\frac{ \mu_\alpha^{ge} \mu_\beta^{eg} \omega_{eg}^3 
	}{ 2 \pi \varepsilon_0 c^3}
    \int_0^\infty d\overline{\xi} \, 
    \int_0^\infty \frac{\overline{k} \, d\overline{k}}{\overline{\kappa}} \,
    I(\overline{\kappa} , x  , \eta) \,
    M_{\alpha\beta} (\overline{k},\overline{\xi})
\nn
&&\times
    \omega_{eg} \, 
    \bigl\{
    \alpha(\overline{\xi}\omega_{eg},1)
    +
    \alpha^{(rc)}(\overline{\xi} \omega_{eg},1,\overline{k}\omega_{eg}) 
    \bigr\}
\label{eqn:T_single_atom}
\;,
\end{eqnarray}
%
%\notes{10a_108}
%
where the energy scale is set by the natural linewidth
$\gamma_{eg} =  |\bfmu^{ge}|^2\omega_{eg}^3 / 
3 \pi \varepsilon_0 c^3$, 
the so-called Lamb-Dicke parameter
$\eta = \omega_{eg} a_0 / c$ gives the size of the trap ground state
in units of the resonant wavelength. The quantity $I$ becomes
\begin{equation}
I(\kappa, x, \eta) 
    \equiv 
    \frac 1 2 \,
    \operatorname{exp}[-2 \overline{\kappa} x + {\overline{\kappa}}^2 \eta^2 ]\,
    (1 + \operatorname{erf}[\frac x {\eta} - \overline{\kappa} \eta ])
\label{eqn:I_def}
\end{equation}
The matrix $M_{\alpha\beta}$ defined in \pr{eqn:M_def}
depends on the reflection coefficients $R^p$ and $R^s$ and encodes
the surface properties. 
% the energy shift can be calculated for different surface models by evaluating
% \pr{eqn:T_single_atom} with the appropriate reflection 
%recoil correction 
%%
%\begin{equation}
%T^{(2)\,rc}_{1 1}
%    =
%    \biggl(
%    \frac 3 4 
%    + \frac {\overline{k}^2 } {2 (a_0 \omega_{eg})^2} 
%    \biggr) \,
%    \frac {1 - \overline{\xi}^2} {1 + \overline{\xi}^2}
%\label{eqn:recoil_single_atom}
%\;.
%\end{equation}
%
%The asymptotic behaviour of the expression
%for large values of $\kappa$ is dealt with in detail in \pr{app:erf}.
%As we show there, the numerical evaluation of 
%the $d\kappa$-integration in \pr{eqn:T_single_atom} can be
%cut off at $\kappa = (d/a_0)^2/x$, 
%the error being $\landau(\operatorname{exp}[-(d/a_0)^2])$.
%This large momentum cut-off corrects for the non-vanishing of the atom 
%wave function \pr{eqn:wfct} at the surface.
%
In the dimensionless units of \pr{eqn:T_single_atom}, the recoil correction $\alpha^{(rc)}$  is now seen to be proportional to
the ratio $\nu / \omega_{eg}$: 
\begin{eqnarray}
\omega_{eg} \, \alpha(\overline{\xi}\omega_{eg},1)
	&=&
	(1+\overline{\xi}^2)^{-1}
\;,
\\
\omega_{eg} \, \alpha^{(rc)}(\overline{\xi}\omega_{eg},1,\overline{k}\omega_{eg})
	&=&
	- \frac {\nu}{\omega_{eg}} \,
	\frac{1-\overline{\xi}^2}{(1+\overline{\xi}^2)^2} \,
	\bigl(
	\frac 3 4 + \frac{\overline{k}^2}{2} \eta^2
	\bigr)
%\label{eqn:}
\end{eqnarray}
The trapping frequency $\nu/2\pi$ for a single ground state atom
in the potential \pr{eqn:V_def} is usually around $10\ldots1000$~Hz,
much smaller than the frequencies of optical transitions
$\omega_{eg}/2\pi \approx 10^{15}$~Hz.
This justifies the expansion of the recoil shift for small atom momenta 
$\V q$ done in \pr{sec:BEC_har}.
Experimental situations where the recoil correction is enhanced in magnitude
could involve tight traps like
optical lattices ($\nu/2\pi \sim 100\,{\rm kHz}$) and Rydberg atoms 
whose transition frequencies can 
be a factor $10^{6}$ smaller \cite{Gallagher}.

The expression in \pr{eqn:T_single_atom} is easily evaluated numerically.
To properly eliminate the exponentially small but nonvanishing overlap of $\phi_0(\V r)$ with the surface,
we cut off the $k$ and $\xi$ integrations at suitably large values,
% off large values of $k$ and $\xi$ in \pr{eqn:T_single_atom} that are 
% responsible for atom-surface
% interactions at very small distances.
as explained in detail in \pr{app:erf}. This procedure applies in exactly 
the same fashion to the integration in \pr{eqn:T_finite_temp} below.

Figure \ref{fig:gold_mirror_recoil} shows the energy shift of a rubidium atom 
in the harmonic trapping potential \pr{eqn:V_def} with $\nu/2 \pi = 1$~kHz. 
At this frequency, the oscillator length is $a_0 \approx 340 \, {\rm nm}$.
As the overlap of the atom wavefunction with the surface should be negligible, 
we restrict the evaluation 
to a distance range $d \geq 2 \, \mu m$, making an error of the
order $\operatorname{Exp}[-(d/a_0)^2] \sim 10^{-16}$.
The black lines in \pr{fig:gold_mirror_recoil} are for the case of a perfectly reflecting surface, 
with the reflection amplitudes $R^p = 1$ and $R^s = -1$.
The red lines involve a frequency-dependent reflection, as appropriate
for a gold surface (described by the Drude model, see \pr{app:gr} for details).
The two terms of \pr{eqn:T_single_atom} are shown separately, the recoil correction (dashed lines) is multiplied by 
a factor of $-\omega_{eg}/\nu$ to fit on the scale.
The dashed horizontal line shows the asymptotic expression for the Casimir-Polder potential 
of an atom in front of a perfect mirror,
\begin{equation}
\Delta E_{CP}(d)
	=
	- \frac{c \, \alpha(0) }{\epsilon_0 d^4} 
	\frac{3}{8 \pi}
\;,
\label{eqn:CP_asymptote}
\end{equation}
with the static polarizability $\alpha(0)$.
%
%\todo{Give formula for Casimir-Polder asymptote and add a horizontal
%line in the plot.}
%
%
%
%
%
%%% FIGURE %%%% FIGURE %%%% FIGURE %%%% FIGURE %%%% FIGURE %%% FIGURE %%
\begin{figure}
\centering
    \includegraphics[width=0.8\textwidth]{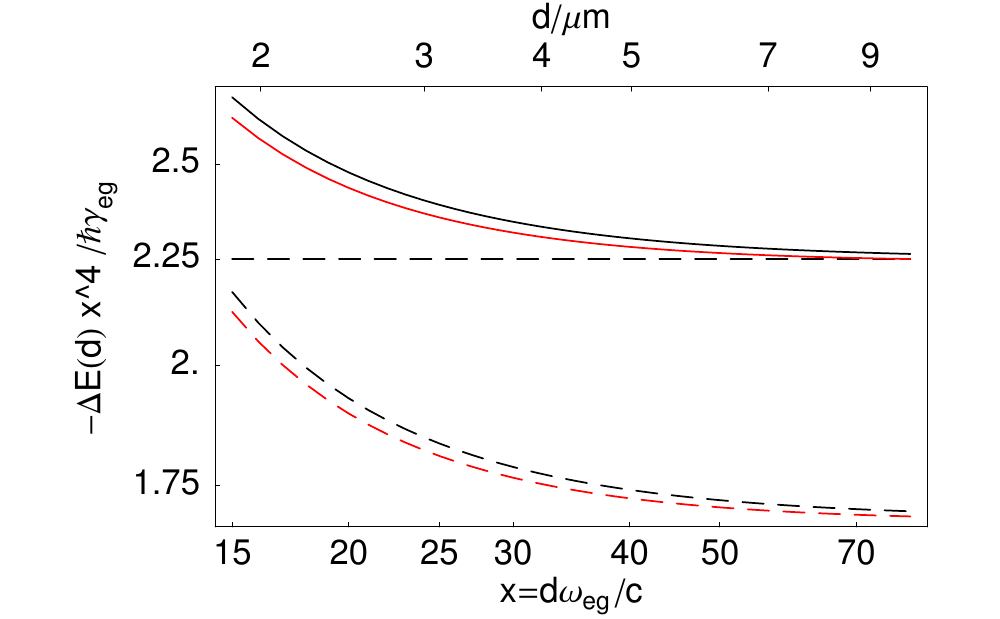}
\caption{
    Atom-surface interaction energy 
    between a rubidium atom  and a  
    perfectly reflecting (black/upper solid) and gold surface (red/lower solid),
    field at zero temperature.
    %\todo{Specify diagonal weights of the polarizability tensor.}
    Distance $d = x c/\omega_{eg}$ in units of the resonance wavelength,
    energy multiplied by $x^4$ and
scaled by the natural linewidth
$\hbar\gamma_{eg} 
= |\bfmu^{eg}|^2 \omega_{eg}^3/(3 \pi \varepsilon_0 c^3)$.
The atom is trapped in the harmonic potential \pr{eqn:V_def} with a trap 
frequency $\nu/2\pi = 1$~kHz.
    Dashed lines: recoil correction multiplied by $- \omega_{eg}/\nu$
    (see \pr{eqn:T_single_atom}).
Resonance frequency 
$\omega_{eg} = 2\pi \, 3.85 \times 10^{14}$~Hz
(isotropic polarizability);
%\notes{10b_18}
parameters of the Drude dielectric function for gold,
\pr{eqn:eps_drude}:
$\omega_p = 5.74\,\omega_{eg}$ and 
$\omega_p \tau = 5 \times 10^3$.
    }
\label{fig:gold_mirror_recoil}
\end{figure}
%%% FIGURE %%%% FIGURE %%%% FIGURE %%%% FIGURE %%%% FIGURE %%% FIGURE %%
%
%
%
%
%
%
%
%
%
%
%%% FIGURE %%%% FIGURE %%%% FIGURE %%%% FIGURE %%%% FIGURE %%% FIGURE %%
\begin{figure}
\centering
	\includegraphics[width=0.8\textwidth]{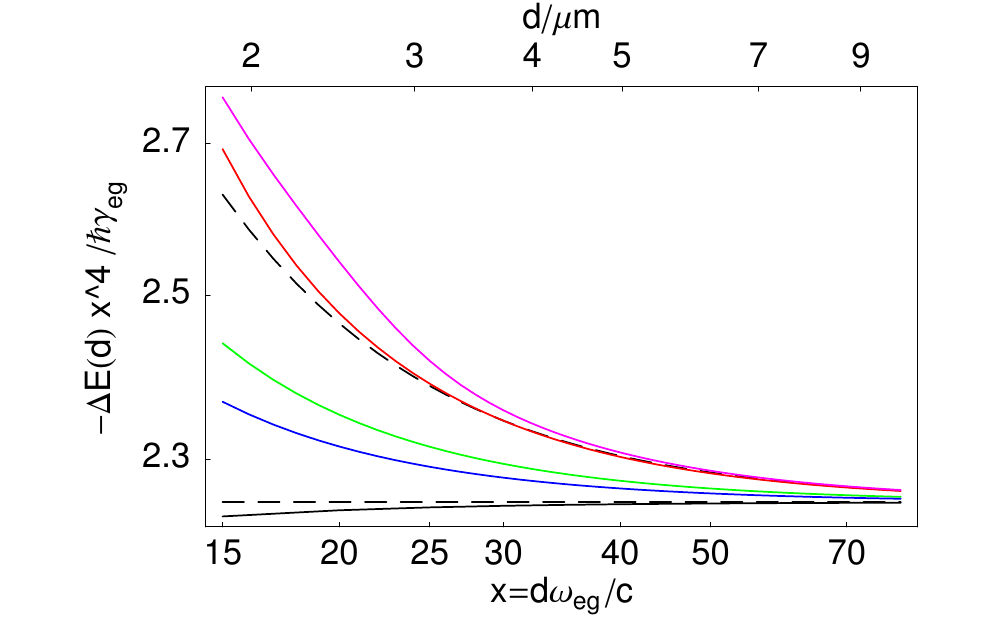}
\caption{
	Atom-surface interaction between a trapped rubidium atom 
	and a perfectly reflecting surface (field at zero temperature).
	Energy scaled by $d^4$ as in \pr{fig:gold_mirror_recoil}.
The full curves correspond to
a perfectly localized atom (black/bottom)
and delocalized atoms with
different trapping frequencies 
($\nu / 2 \pi  = \; 3, 2, 1$~kHz in blue, green, red from bottom to top).
	Top/pink curve: interaction energy per atom for a trapped ideal 
	Bose gas at $T=0.2 \, T_c$ and $\nu / 2\pi = 1$~kHz 
	(see \pr{eqn:T_finite_temp}).
	Horizontal dashed line: asymptotic expression $\Delta E_{CP}$ (see \pr{eqn:CP_asymptote}) 
	for the Casimir-Polder potential.
	Dashed curve: $\Delta E_{CP}$ multiplied with the enhancement factor
	\pr{eqn:estimate-curvature-average} for $\nu/2\pi = 1$~kHz. 
}
\label{fig:trapped__no_recoil}
\end{figure}
%%% FIGURE %%%% FIGURE %%%% FIGURE %%%% FIGURE %%%% FIGURE %%% FIGURE %%
%
%
%
%
%
%
%
%
%
%
%
%
%
%
%
%
%
%
%
%
%
%
%
%
%%%%%%%%%%%%%%%%%%%%%%%%%%%%%%%%%%%%%%%%%%%%%%%%%%%%%%%%%%%%%%%%%%%%%%%%%%%%%%%%%%%%%%%%%%%%%%%%%%%%
\section{Ideal {B}ose gas in a surface trap}
\label{sec:ideal}
%%%%%%%%%%%%%%%%%%%%%%%%%%%%%%%%%%%%%%%%%%%%%%%%%%%%%%%%%%%%%%%%%%%%%%%%%%%%%%%%%%%%%%%%%%%%%%%%%%%%
%
Now we will use the results obtained in \pr{subsubsec:atom_part_ideal} to calculate the atom-surface 
interaction for a gas of $N$ noninteracting bosons.  
The atomic system (treated in the grand-canonical ensemble) is supposed to be in thermal equilibrium 
at the inverse temperature $\beta$, but its temperature is allowed to be different from the
field temperature.

\subsection{Atomic correlation function}

As above in \pr{sec:BEC_har}, we will assume an isotopic harmonic trapping potential of the form \pr{eqn:V_def}.  
For the ideal Bose gas, $\phi_{\V n}$ and $E_{\V n}$ that enter the expression 
\pr{eqn:Psi_g_ideal} for  $\Psi_g$ are then simply the eigenfunctions and energies 
of a three-dimensional  harmonic oscillator:
\begin{equation}
E_{\V n}
	=
	(n_x + n_y + n_z) \, \nu,
	\quad
	n_i = 0, 1, 2, \dots
\;,	
%\label{eqn:}
\end{equation}
where the ground state of the trap has been set equal to the zero of energy.
The critical temperature takes the value~%
\cite{Stringari,Barnett_2000}%\cite[eqn.~(4.21)]{Barnett_2000}.
\begin{equation}
T_c 
	=
	\nu \, 
	\biggl(
	\frac{N}{\zeta[3]}
	\biggr)^{1/3}
	.
%\label{eqn:}
\end{equation}
For a given mean particle number $N$ and inverse temperature $\beta$, 
the (negative valued) chemical  potential $\mu(N,\beta)$ has  to be determined from the relation 
\begin{equation}
N(\mu,\beta) 
	=
	\int d^3 r \,
	\langle \Psi^\dagger_g(\V r) \Psi_g(\V r) \rangle
\;,
\label{eqn:ideal_number}
\end{equation}
where the brackets $\langle \dots \rangle$ denote a state of the atomic system at temperature $\beta$.

With the same arguments as in \pr{subsec:recoil-correction} above, 
the propagator for the excited atomic state is local to a very good approximation.
Neglecting the small recoil correction, we find from 
\pr{eqn:atom_part_ideal}
\begin{eqnarray}
%\AtomPartIdeal{-0.5ex}{0.2\textwidth}
\langle \Psi_g^\dagger (x_2) \Psi_g (x_1) \rangle \,
\contraction{}{\Psi_e}{(x_2)}{\Psi_e^\dagger}
\Psi_e(x_2)\Psi_e^\dagger(x_1)
	&\approx&
	\sum_{\V n} \, 
	\frac{\Phi^*_{\V n} (\V{r}_2) \Phi_{\V n} (\V{r}_1)}{\operatorname{exp}[\beta(E_{\V n} -\mu)] -1}
	\operatorname{exp}[-i (t_2 - t_1) (\omega_{eg} - E_{\V n})]
\nonumber
\\
	&&\times
	\Theta(t_2 - t_1) \,
	\delta(\V{r}_2 - \V{r}_1) \,
\;.
\label{eqn:atom_ideal_approx1}
\end{eqnarray}
%\notes{10b_51}
To the same precision, we can neglect the 
% In order to simplify \pr{eqn:atom_ideal_approx1} further, we note that  for low 
% temperatures the 
single particle energy $E_{\V n}$ compared to 
the atomic transition energy $\omega_{eg}$. 
%At a temperature $T =  (T/T_c) \, T_c$, the mean thermal energy is 
%$T = (T/T_c) (N/\zeta[3])^{1/3} \nu$.
This is even true for realistic atom temperatures:
for a trapping frequency $\nu/2\pi = 1$~kHz and a mean number of $N=10^4$ trapped particles,
the mean thermal energy that sets the scale for the relevant $E_{\V n}$
evaluates to 
$T = (T/T_c) \, 2\pi~20.3$~kHz much smaller than 
$\omega_{eg}$. % = 2\pi~3.84\times 10^{11}$~kHz for rubidium.
We thus neglect $E_{\V n}$ in the exponential 
in \pr{eqn:atom_ideal_approx1} 
and obtain
\begin{eqnarray}
\AtomPartIdeal{-0.5ex}{0.15\textwidth}
	&\approx&
	\Theta(t_2 - t_1) \,
	\delta(\V{r}_2 - \V{r}_1) \,
	\langle
	\Psi_g^\dagger (\V{r}_2) \Psi_g(\V{r}_1) \,
	\rangle
	e^{-i \omega_{eg} (t_2 - t_1)}
\;.
\label{eqn:atom_ideal_approx2}
\end{eqnarray}
% With the rescaled chemical potential 
% $\overline\mu = \mu / \nu$
% and inverse temperature
% $\overline\beta = \beta \nu$,
The correlation function
$\langle \Psi^\dagger_g(\V{r}_2) \Psi_g(\V{r}_1) \rangle$ that enters in
\pr{eqn:ideal_number}  and in \pr{eqn:atom_ideal_approx2} above 
reads~\cite{Stringari,Barnett_2000}
\begin{eqnarray}
\langle \Psi_g^\dagger (x_2) \Psi_g(x_1) \rangle
	&=&
	\bigl( \sqrt{\pi} a_0  \bigr)^{-3}
	\sum_{j=1}^\infty\biggl\{
	e^{j \beta \mu}\,
	\bigl( 1 - e^{-2 j \beta \nu }  \bigr)^{-3/2} \,
\label{eqn:corr1}
\\	&&	
	{}
	\times \operatorname{exp} 
	\biggl[
	- \frac{1}{4 a_0^2} \, 
	\bigl(
	|\V {r}_2^d + \V {r}_1^d|^2 \,
	\operatorname{tanh[j \beta \nu /2]}
	+
	|\V {r}_2^d - \V {r}_1^d|^2 \,
	\operatorname{coth[j \beta \nu /2]}
	\bigr)
	\biggr]
	\biggr\}
\nonumber
\end{eqnarray}
%
%\notes{09f_101}
%(see \cite[eqn.~(4.15)]{Barnett_2000}),
where the vectors $\V{r}^d \equiv (\V x, z-d)$ account for the distance $d$ between
the surface and the center of the trap.

%\material{
%Recoil expansion for the atomic propagator \pr{eqn:atom_part_ideal}: 
%%
%\begin{align}
%\int \frac{d^3 q}{(2 \pi)^3} &\,
%e^{i \V{q}.(\V{x_2} - \V{x_1})} \,
%e^{-i (q^2/(2m) + \omega_{eg}) (t_2 - t_1)} 
%\\
%	&=
%	\int \frac{d^3 q}{(2 \pi)^3} \,
%	e^{i \V{q}.(\V{x_2} - \V{x_1})} \,
%	(1 - i \frac{q^2}{2m} (t_2 - t_1) + \landau(q^4))\,
%	e^{-i  \omega_{eg} (t_2 - t_1)} 
%\\	
%	&\approx
%	\delta(\V{x}_2 -\V{x}_1)\,
%	e^{-i \omega_{eg} (t_2-t_1)}
%\\
%	&-
%	i \int \frac{d^3 q}{(2\pi)^3} \, 
%	\operatorname{exp}[i \V q . ({\V{x}_2} - \V{x}_1)]\,
%	\operatorname{exp}[-i \omega_{eg}(t_2 - t_1)]\,
%	\frac{q^2}{2 m} (t_2 -t_1)
%\;.
%%\label{eqn:}
%\end{align}
%%
%\notes{10a_84}
%}
%which is valid as long as the kinetic energy of the excited atom (the recoil) is small compared to the 
%atomic transition energy $\omega_{eg}$.
%We will use the approximate expression
%%
%\begin{eqnarray}
%\langle 
%\contraction{} {\Psi} {(x_2)} {\Psi}
%\Psi(x_2) \Psi^\dagger(x_1)
%\rangle \,
%	\approx	
%	\delta(\V{x}_2 - \V{x}_1)
%	\Theta(t_2 - t_1) \,
%	e^{-i  \omega_{eg}) (t_2 - t_1)} \,
%	G^{(1)} (\V{x}_2,\V{x}_1)
%%\label{eqn:}
%\end{eqnarray}
%%

\subsection{Surface-induced energy shift}

With these approximations,
the general expression \pr{eqn:S2N0b} 
% together with the atomic part \pr{eqn:atom_part_ideal}
% to calculate the attractive potential between a trapped ideal Bose gas and a 
% surface.
% With this we get for the 
gives a $T$-matrix
%\notes{10b_52}
%
\begin{equation}
\langle N | T^{(2)} | N \rangle
	=
	\mu_\alpha \mu_\beta \,
	\int d^3 r \,
	\langle
	\Psi_g^\dagger (\V{r}) \Psi_g(\V{r}) \,
	\rangle
	\int \frac{d\omega}{2\pi} \,
	\frac{\tilde{D}^F_{\alpha\beta} (\V{r},\V{r},\omega)} {\omega - \omega_{eg} + i\epsilon}
%\label{eqn:}
\;.
\end{equation}
Had we kept the trap eigenenergy $E_{\V n}$, it would appear as a 
small shift of $\omega_{eg}$ in the denominator.
Using \pr{eqn:f2rN0} and neglecting  any thermal photons 
(see the remark below \pr{eqn:f2rN0}),
we obtain
\begin{equation}
\langle N | T^{(2)} | N \rangle
	=
	-\mu_\alpha \mu_\beta \frac 1 \pi \,
	\int d^3 r \,
	\langle
	\Psi_g^\dagger (\V{r}) \Psi_g(\V{r}) \,
	\rangle
	\int_0^\infty d\xi  \,
	\tilde{G}(\V{r},\V{r},i\xi) \,
	\frac{\omega_{eg}}{\omega_{eg}^2 + \xi^2}
\;.
%\label{eqn:}
\end{equation}
Performing the spatial integration and switching to the dimensionless
variables of \pr{eqn:T_single_atom} yields
\begin{eqnarray}
\langle N | T^{(2)} | N \rangle
    &=& 
    -\frac {1} {2 \pi} \, 
    \frac{ \mu_\alpha^{ge} \mu_\beta^{eg} \omega_{eg}^3 }{ 
    \varepsilon_0 c^3 }\,
    \sum_{j=1}^\infty \,
    e^{j \beta \mu } \,
    \bigl(
    (1 - e^{-2 j \beta \nu } ) \, \operatorname{tanh}[j \beta \nu /2]
    \bigr)^{-3/2}
\nonumber
\\
&&\times
    \int_0^\infty d\overline{\xi} \, 
    \int_0^\infty \frac{\overline{k} \, d\overline{k}}{\overline{\kappa}} \,
    I(\overline{\kappa} , x  , \eta_+ ) \,
    M_{\alpha\beta} (\overline{k},\overline{\xi},R^p , R^s) \,
    \frac 1 {1 + \overline{\xi}^2}
\label{eqn:T_finite_temp}
\;,
\end{eqnarray}
where $I(\overline{\kappa} , x  , \eta_+ )$ is defined in  \pr{eqn:I_def},
and the matrix $M_{\alpha\beta}$ in \pr{eqn:M_def}.
The Lamb-Dicke parameter 
$\eta_+ = a_+ \omega_{eg} / c$ now involves 
the temperature dependent width
\begin{equation}
a_+
	=
	a_0 \,
	(\operatorname{tanh}[j \beta \nu  /2])^{-1/2}
	\geq a_0
\;.
\label{eqn:aplus_def}
\end{equation}
To compare \pr{eqn:T_finite_temp} with the 
result \pr{eqn:T_single_atom} for the single atom,
we note that the constraint \pr{eqn:ideal_number} leads to
\begin{equation}
\sum_{j=1}^\infty \,
e^{j \beta \mu } \,
\bigl(
(1 - e^{-2 j \beta \nu } ) \, \operatorname{tanh}[j \beta \nu /2]
\bigr)^{-3/2}
	=
	N
%\label{eqn:}
\end{equation}
and consider an interaction energy per atom, 
$\langle N | T^{(2)} | N \rangle / N$. The terms with large $j$ in the sum
involve a width $a_+$ equal to the zero-temperature value 
$a_0$. These terms describe the condensate atoms in the trap ground state.
The terms with small $j$ have larger values of $a_+$ and contribute to 
the energy shift as a broader trap would do. Indeed, for $j = 1$
and $\beta \nu  \ll 1$, one gets the spatial width of a classical,
thermal density distribution. 

This behaviour is shown in the numerical evaluation 
of \pr{eqn:T_finite_temp} and \pr{eqn:T_single_atom}
in \pr{fig:trapped__no_recoil}, for a perfectly reflecting surface. 
(More realistic materials can be described without further complications.)
The atom-surface interaction per atom at $T = 0.2 \, T_c$ is larger than 
for a single atom (at the same trap frequency $\nu / 2\pi = 1$~kHz), which
is due to the larger spatial size of the thermally excited trap levels.

At an atom-surface distance of $d > 2\mu m$, the interaction potential for the perfectly localized atom 
(calculated from \pr{eqn:cc_wylie}) is already deep in the retarded $x^{-4}$ regime.
For an atom delocalized in the trap, the interaction potential becomes larger 
in magnitude because of the curvature of the
Casimir-Polder interaction. Averaging a power law $1/z^4$ over a narrow
distribution ($\sigma \ll d$) centered at $z = d$, we get to leading order
the enhancement factor
\begin{equation}
	\big\langle \frac{ 1 }{  z^4 } \big\rangle \approx 
	\frac{ 1 }{  d^4 } \left[ 1 + 
	5 (\sigma / d)^2  + \ldots \right]
	\label{eqn:estimate-curvature-average}
\;.
\end{equation}
The dashed black curve in \pr{fig:trapped__no_recoil} shows the asymptotic expression for the Casimir-Polder potential
\pr{eqn:CP_asymptote} multiplied with the above enhancement factor for a trapping frequency of $\nu/2\pi = 1$~kHz.
The estimate \pr{eqn:estimate-curvature-average} is seen to be in good agreement with our result from \pr{eqn:T_single_atom} (red line).
% the trapped atom has a finite probability of being found at a distance smaller 
% than $d$ to the surface,
% which enhances the Casimir-Polder interaction.

%The correlation-function looks very much like the integal of our original Gaussian wavefunction,
%but now the width is temperature dependent:
%%
%\begin{equation}
%G^{(1)} (\V{x},\V{x})
%	=
%	(\sqrt{\pi} a_0)^{-3} \,
%	\sum_{j=1}^\infty \,
%	\frac{\operatorname{exp}[j \overline \beta \overline \mu]}{(1 - \operatorname{exp}[-2 j \overline \beta])^{3/2}} \,
%	\operatorname{exp}[- \frac{1}{a_+^2} (x^2 + y^2 + (z-d)^2)]
%%\label{eqn:}
%\end{equation}
%%
%Compare
%\notes{10a_86}
%%
%\begin{equation}
%\Phi(\V{x}) \Phi^*(\V{x}) 
%	=
%	(\sqrt{\pi} \sigma)^{-3} \,
%	\operatorname{exp}[- \frac{1}{\sigma^2} (x^2 + y^2 + (z-d)^2)]
%%\label{eqn:}
%\end{equation}
%
%
%
%
%
%
%
%
%
%
%
%%%%%%%%%%%%%%%%%%%%%%%%%%%%%%%%%%%%%%%%%%%%%%%%%%%%%%%%%%%%%%%%%%%%%%%%%%%%%%
\section{Summary and outlook}
\label{sec:sum}
%%%%%%%%%%%%%%%%%%%%%%%%%%%%%%%%%%%%%%%%%%%%%%%%%%%%%%%%%%%%%%%%%%%%%%%%%%%%%%
%
The starting point of our calculation was a second-quantized Hamiltonian that  
describes the interaction of a trapped system of $N$ atoms  
with the electromagnetic field.
We have focused on two simple models for the atomic system:
an interacting BEC described by $N_0$ atoms populating a single condensate wave function
(described by the state  $|N_0\rangle$) 
and a noninteracting Bose gas at finite temperature, where the $N$ particles populate the
various single particle states of the trap (this state is denoted schematically
by $|N\rangle$).
%on the degenerate limit where $N_0$ atoms populate
%a single condensate wave function, described by the state $| N_0 \rangle$.
To calculate the interaction energy between the atoms and a plane surface,
we made a perturbative expansion of the electromagnetic self-energy
and worked out the $T$-matrix elements 
$\langle N_0 | T^{(2)} | N_0 \rangle$ 
and
$\langle N | T^{(2)} | N \rangle$ 
to second order in the atom-field coupling. The methods developed here
are general enough to push the diagrammatic expansion to higher
orders.
The electric field propagator has been expressed in terms of retarded 
Green functions that permit to identify easily the contribution brought
about by the surface.
The characteristics of the surface material then enter 
through the scattering amplitudes for light, 
which allows for treating a wide range  of materials.
For the sake of simplicity, we considered the field to be at zero temperature
as well,
but thermal corrections can be included in a straightforward way by considering
the temperature dependent term in \pr{eqn:f2rN0}. Even non-equilibrium
situations (bodies at different temperatures) can be covered by combining 
the techniques of fluctuation electrodynamics \cite{Rytov3} with 
the Keldysh formalism (see Ref.\cite{Mkrtchian09a} for an example).

The expression found for $\langle N_0 | T^{(2)} | N_0 \rangle$ in \pr{eqn:TN0.1} 
describes the Casimir-Polder like interaction energy  
of a trapped Bose gas with the surface,
for a general condensate wave function $\phi_0(\V x)$.
%We worked out the detailed form of $T^{(2)}_{N_0 N_0}$  for a Gaussian profile 
%of the ground state
%wave function in \pr{eqn:TN0.3}.
If the system is reduced to a perfectly localized single atom as treated in \cite{Wylie_1985}, 
our expression reproduces known results  (see \pr{eqn:cc_wylie}).
It also highlights that in full generality, the atom-surface interaction does not
reduce to an integral over the density distribution of the atoms, due to
the (virtual) propagation in the excited state.
The Bose gas-surface interaction energy shows an overall scaling with the 
atom number $N_0$ 
(as can be expected at this order of perturbation theory),
but even the interaction energy per atom still depends weakly on $N_0$. 
We have identified for this dependence the following physical mechanisms.
(i) The interaction energy involves a spatial average over the density 
profile whose width is larger for repulsive atom-atom interaction. This
effect was already taken into account in the pioneering experiments 
of Ref.\cite{Harber_2005, Obrecht_2007}.
(ii) The atomic interactions (treated here as a contact potential) 
shift the optical transition frequency (see for example the experiments of
Ref.\cite{Heinzen00}) and modifies the
ground-state polarizability \pr{eqn:alpha_def}.
(iii) The optical spectral line is recoil-broadened due to the kinetic 
energy of the atoms. This effect is very weak for typical traps and in
the fully degenerate limit as the phase gradient of the condensate wave 
function vanishes.

For the ideal Bose gas from 
$\langle N | T^{(2)} | N \rangle$ in  \pr{eqn:T_finite_temp},
the Casimir-Polder interaction per particle 
does not depend on the atom number.
We showed that the influence of a higher atom temperature on the atom-surface interaction is similar to that of a broadening of the
trap potential.

We plan to generalize the method presented in the present paper in 
two directions:
on the BEC side of the problem, we want to include contributions from higher collective modes (condensate depletion, phase fluctuations, thermal 
density fraction) and revisit the problem of two atoms in front of a 
surface \cite{Messina_2008}. This setting has also been realized in a many-body
version, by splitting a BEC in two spatially separated 
modes~\cite{Gati06,Hofferberth07}. 
On the cavity QED side, higher orders of atom-photon interactions will
be considered where intensity fluctuations of the quantum field
near the surface appear~\cite{Ford02, Novotny04a}.
\begin{appendix}
%
%
%%%%%%%%%%%%%%%%%%%%%%%%%%%%%%%%%%%%%%%%%%%%%%%%%%%%%%%%%%%%%%%%%%%%%%%%%%%%%%
\section{Retarded Green function for the electric field in the presence of
	 an interface}
\label{app:gr}
%%%%%%%%%%%%%%%%%%%%%%%%%%%%%%%%%%%%%%%%%%%%%%%%%%%%%%%%%%%%%%%%%%%%%%%%%%%%%%
%
The reflected part of the retarded Green function in the presence of an interface
(as presented in \cite[eqn.~(3.4)]{Wylie_1984}, see also \cite[sec.~2]{Panasyuk_2009} for an overview) 
reads 
%
%\todo{$\bullet$ check for $\varepsilon_0$, sorry!}
%
\begin{equation}
G^R_{\alpha\beta}(\V {r}_1, \V{r}_2, \omega) 
	= 
	- \frac{i \omega^2}{2\pi \varepsilon_0 c^2} \, 
	\int \frac{ {\rm d}^2 k}{ k_z } \,
	R_{\alpha\beta}(\V k , \omega) \,
	e^{	i k_z (z_1+z_2)  
		+ i \V k . (\V{x}_1 - \V{x}_2)}
\;,
\label{eqn:G_def_app}
\end{equation}
with
$
k_z = \sqrt{\omega^2/c^2 - {k}^2}.
$
Here, the two-dimensional vectors
$\V x$ and $\V k$ denote the position and momentum vectors parallel to 
the surface, respectively.
Henceforward in this appendix, we use units with $c = \varepsilon_0 =1$.
The matrix $R_{\alpha\beta}(\V k , \omega)$ is defined as
\begin{equation}
R_{\alpha\beta}(\V k , \omega)
	=
	(\hat s \hat s)_{\alpha\beta} \, R^s  
	+ (\hat{p}_{0+} \hat{p}_{0-})_{\alpha\beta} \, R^p
\label{eqn:R_def_app}
\end{equation}
The functions $R^s$ and $R^p$ in \pr{eqn:R_def_app}
are the Fresnel reflection coefficients for $s$- and $p$-polarized light,
which can be modeled to realize different surface materials.
For the case of a perfectly reflecting surface,
$R^s = -1$ and $R^p = 1$,
while in general the reflection coefficients are frequency dependent
(see \cite{Sipe_1981,Wylie_1984, Wylie_1985}):
Considering an interface between vacuum ($\eps_0 = 1$) and a material with
a local and isotropic dielectric function $\eps(\omega)$,
$R^s$ and $R^p$ are given by
\begin{eqnarray}
R^s 
	&=&
	\frac	{k_z - (\omega^2\, \eps( \omega ) - k^2)^{1/2}}
		{k_z + (\omega^2\, \eps( \omega ) - k^2)^{1/2}}
	\label{eqn:Rs_def}
	\;,
\\
R^p
	&=&
	\frac	{\eps \, k_z - (\omega^2\, \eps( \omega ) - k^2)^{1/2}}
		{\eps \, k_z + (\omega^2\, \eps( \omega ) - k^2)^{1/2}}
	\label{eqn:Rp_def}	
	\;.
\end{eqnarray}
In \pr{subsec:single_atom}, we use the Drude model for a metal surface, with
\begin{equation}
\eps(\omega)
	=
	1 
	- \frac {\omega_p^2} {\omega (\omega + i/\tau)}
\label{eqn:eps_drude}
\;,
\end{equation}
where $\omega_p$ is the plasma frequency and $\tau$ the collision time.
Finally, the dyadic elements 	$(\hat s \hat s)_{\alpha\beta}$ and 
$(\hat{p}_{0+} \hat{p}_{0-})_{\alpha\beta}$ in \pr{eqn:R_def_app} 
involve the normalized polarization vectors
\begin{eqnarray}
\hat {\V{s}} 	
	&=& 
	\hat{\V k} \times \hat{\V z}
\\
\hat{\V{p}}_{0 \pm} 
	&=& 
	\frac{ k \,\hat{\V z} \mp k_z \hat{\V k} }{\omega}
\;.
\end{eqnarray}
%
%
%
%
%
%
%
%
%
%
%
%
%
%
%
%
%
%
%
%
%%%%%%%%%%%%%%%%%%%%%%%%%%%%%%%%%%%%%%%%%%%%%%%%%%%%%%%%%%%%%%%%%%%%%%%%%%%%%%
\section{Approximating the error function integral}
\label{app:erf}
%%%%%%%%%%%%%%%%%%%%%%%%%%%%%%%%%%%%%%%%%%%%%%%%%%%%%%%%%%%%%%%%%%%%%%%%%%%%%%
%
In the integrands of \pr{eqn:T_single_atom} and \pr{eqn:T_finite_temp}, we encounter 
the expression
\begin{equation}
I(\overline{\kappa}, x, \eta ) 
    \equiv 
    \frac 1 2 \,
    \operatorname{exp}[-2 \overline{\kappa} x + {\overline{\kappa}}^2 \eta^2 ]\,
    (1 + \operatorname{erf}[\frac x {\eta} - \overline{\kappa} \eta ])
\;
\end{equation}
where $\overline{\kappa}$ is integrated from zero to infinity, $\eta = 
a_0 \omega_{eg} / c$ is fixed by 
the atomic transition frequency and mass and the trap geometry and the 
positive distance $x$ varies such that
$x > \eta^2$ is always fulfilled.

Noting that the argument of the error function changes sign at 
$\overline{\kappa} =  x/\eta$,
we can approximate the error function for large values of $\overline\kappa$ 
(see \cite[eqn.~(8.254)]{Gradshteyn}) to obtain
\begin{eqnarray}
I(\overline{\kappa}, x, \eta )  	
	&\approx&  
	\frac{ 
	\operatorname{exp}[-x^2/\eta^2]
	}{
	2 \sqrt \pi \,  ( \overline\kappa \eta - x/\eta )
	}
\;,
\quad \text{for} \; 
\overline\kappa \gg \frac{x}{\eta^2}
\label{eqn:cut_off}
\end{eqnarray}
which is exponentially small in the quantity $(x/\eta)^2$.
%
% \begin{eqnarray}
% I(\overline{\kappa}, x, a_0 \omega_{eg})  	
%	&=&
%	\landau\bigl( \operatorname{exp}[-(x^2/\eta^2] \bigr)
%\;,
%\;.
%\end{eqnarray}
%
In numerical integrations, we will thus cut off the $d \overline \kappa$-integration at 
$\overline \kappa= x / \eta^2$,
omitting terms of order $\landau\bigl( \operatorname{exp}[-(x/\eta)^2] \bigr)$ in the integrand.
The neglected quantities are small: for a rubidium atom at $T=0$ 
trapped in a $\nu / 2\pi = 1$~kHz trap at an atom-surface distance 
of $x = d \omega_{eg} / c = 15$, we have $(x/\eta)^2 \approx 30$.
Conceptionally, the high momentum cut-off is necessary as the atomic probability density 
$|\phi_0( {\bf r} ) |^2$ we adopt here
is not zero at the surface, but only exponentially small,
namely of the same order as the terms neglected 
in \pr{eqn:cut_off}.

%where $\myw  = \sqrt{1 - k^2 }$.
%Setting $\myw  = i \kappa$  for $k > 1$ ($\kappa \ge 0$),
%\pr{eqn:erf_1} reads
%%
%\begin{equation}
%I(k, x, d/a_0) 
%	= 
%	\frac 1 2 e^{- 2 \kappa x } \, e^{+ [\kappa x a_0 / d ]^2} 
%	(1 + \operatorname{erf}[\frac d {a_0} - \kappa\frac{x}{d/a_0} ])
%\label{eqn:I_kappa_def}
%\;.
%\end{equation}
%%
%%(This is the expression encountered in \pr{eqn:Tgg_1}, where we had to
%%consider only the domain $\kappa \ge 0$.)
%Noting that the argument of the error function changes sign at 
%$\kappa = (d/a_0)^2 / x \gg 1$,
%we can approximate the error function for large values of $\kappa$ to obtain
%%
%\begin{equation}
%I(\kappa, x, d/a_0)  	\approx  
%	\frac 	{e^{- (d / a_0)^2}}
%		{2 \sqrt \pi \,  (\kappa \frac{x}{d/a_0}  - \frac d {a_0})}
%	=
%	\landau\bigl( \operatorname{exp}[-(d/a_0)^2] \bigr)
%\;,
%\quad \text{for} \; 
%\kappa \gg \frac{(d / a_0)^2}{x}
%\;.
%\label{eqn:cut_off}
%\end{equation}
%%
%In numerical integrations, we will thus cut off the $d \kappa$-integration at 
%$\kappa=(d / a_0)^2 / x$,
%omitting terms of order $\landau\bigl( \operatorname{exp}[-(d/a_0)^2] \bigr)$.
%This high momentum cut-off is necessary as the atomic probability density 
%$|\phi_0( {\bf x} ) |^2$ we adopt here,
%is not zero at the surface, but only exponentially small
%(of the same order as the terms neglected 
%in \pr{eqn:cut_off}).
%For a perfectly localized atom as treated in \cite{Wylie_1985}, we have 
%%
%\begin{equation}
%\lim_{a_0 \to 0} 
%	I(k, x, d/a_0)
%	=
%	e^{- 2 \kappa x},
%	\qquad (k > 1)
%\;, 
%\end{equation}
%%
%and the integral over \pr{eqn:I_kappa_def}
%is convergent at large $\kappa$.
%
%
%
%
\end{appendix}
%
%
%
%
%
%
%
%
%
%
%%%%%%%%%%%%%%%%%%%%%%%%%%%%%%%%%%%%%%%%%%%%%%%%%%%%%%%%%%%%%%%%%%%%%%%%%%%%
% References
%%%%%%%%%%%%%%%%%%%%%%%%%%%%%%%%%%%%%%%%%%%%%%%%%%%%%%%%%%%%%%%%%%%%%%%%%%%%  
%
\bibliographystyle{unsrt}
\bibliography{collectiveCP}	   
\end{document}